\documentclass[twocolumn]{aastex63}

\usepackage{amsmath,bm}
\usepackage{graphicx}
\usepackage{enumitem}
\usepackage{tabularx}
\usepackage{csvsimple}

\newcommand{\isofont}[1]{{\mathrm{#1}}}
\newcommand{\isomass}[1]{{\ensuremath{\isofont{^{#1}}}}}
\newcommand{\isocharge}[1]{{\ensuremath{\isofont{_{#1}}}}}
\newcommand{\isotope}[3]{{\ensuremath{\isocharge{#1}\isomass{#2}\isofont{#3}}}}

\newcommand{\I}[2]{{\isotope{}{#1}{#2}}}

\newcommand{\Ep}[1]{{\ensuremath{10^{#1}}}}
\newcommand{\E}[1]{{\ensuremath{\powersep\Ep{#1}}}}

\newcommand{\powersep}{\times}

\newcommand{\unitstyle}[1]{\ensuremath{\mathrm{#1}}}

\newcommand{\Kelvin}{\unitstyle{K}}
\newcommand{\K}{\Kelvin}  

\newcommand{\Msun}{\ensuremath{\unitstyle{M}_\odot}}
\newcommand{\Lsun}{\ensuremath{\unitstyle{L}_{\odot}}}

\newcommand{\code}[1]{\texttt{#1}}
\newcommand{\mesa}{\code{MESA}}
\newcommand{\MESA}{\mesa}

\newcommand{\GYRE}{\code{GYRE}}

\input{macros/vectors.tex}

\newcommand{\D}{{\mathrm d}}

\newcommand{\nuclei}[2]{\ensuremath{\mathrm{^{#1}#2}}}

\newcommand{\COrate}{\ensuremath{\mathrm{^{12}C(\alpha, \gamma)^{16}O}}}

\newcommand{\helium}[1][4]{\nuclei{#1}{He}}

\newcommand{\carbon}[1][12]{\nuclei{#1}{C}}
\newcommand{\nitrogen}[1][14]{\nuclei{#1}{N}}
\newcommand{\oxygen}[1][16]{\nuclei{#1}{O}}

\newcommand{\neon}[1][20]{\nuclei{#1}{Ne}}

\newcommand{\Teff}{\ensuremath{T_{\rm eff}}}	

\newcommand{\BV}{Brunt-V\"{a}is\"{a}l\"{a}}

\newcommand{\revision}[1]{#1}

\newlength{\apjcolwidth}
\newlength{\figwidth}
\newlength{\doublewide}
\newcommand{\tightitems}{
  \setlength{\topsep}{0pt}
  \setlength{\itemsep}{-1pt}
  \setlength{\parsep}{0pt}
  \setlength{\parskip}{0pt}
}

\usepackage{lineno}
\usepackage{soul,xcolor}

\begin{document}
\setstcolor{red}
\title{Seismic Signatures of the $^{12}$C($\alpha, \gamma$)$^{16}$O Reaction Rate in White Dwarf Models with Overshooting}

\shorttitle{Signatures of $^{12}$C($\alpha, \gamma$)$^{16}$O in WD models with  overshooting}
\shortauthors{Chidester, Timmes, \& Farag} 

\author[0000-0002-5107-8639]{Morgan T. Chidester}
\affiliation{School of Earth and Space Exploration, Arizona State University, Tempe, AZ 85287, USA}

\author[0000-0002-0474-159X]{F.X.~Timmes}
\affiliation{School of Earth and Space Exploration, Arizona State University, Tempe, AZ 85287, USA}

\author[0000-0002-5794-4286]{Ebraheem Farag}
\affiliation{School of Earth and Space Exploration, Arizona State University, Tempe, AZ 85287, USA}

\correspondingauthor{Morgan T. Chidester}
\email{taylormorgan32@gmail.com}


\begin{abstract}
We consider the combined effects that overshooting and the \COrate\ reaction rate \revision{have} on variable white dwarf stellar models.  We find that carbon-oxygen white dwarf models continue to yield pulsation signatures of the current experimental \COrate\ reaction rate probability distribution function when overshooting is included in the evolution.  These signatures hold because the resonating mantle region, encompassing $\simeq$\,0.2\,\Msun\ in a typical $\simeq$\,0.6\,\Msun\ white dwarf model, still undergoes radiative helium burning during the evolution to a white dwarf. Our specific models show two potential low-order adiabatic g-modes, $g_2$ and $g_6$, that signalize the \COrate\ reaction rate probability distribution function.  Both g-mode signatures induce average relative period shifts of $\Delta P/P = 0.44 \%$ and $\Delta P/P =  1.33\%$  for $g_2$ and $g_6$ respectively. We find that $g_6$ is a trapped mode, and the $g_2$ period signature is inversely proportional to the \COrate\ reaction rate.  The $g_6$ period signature generally separates the slower and faster reaction rates, and has a maximum relative period shift of $\Delta P/P = 3.45\%$.  We conclude that low-order g-mode periods from carbon-oxygen white dwarfs may still serve as viable probes for the \COrate\ reaction rate probability distribution function when overshooting is included in the evolution. 
\end{abstract}


\section{Introduction}

Helium burning is primarily the fusion of helium into carbon by the triple-alpha (3$\alpha$) process. 
All stars born with more than $\simeq$ 0.5\,\Msun\ go through this stage of energy production as they evolve beyond the main-sequence \citep[e.g., ][]{hansen_2004_aa}.
Helium burning also plays a key role in transients such as  
Type I X-ray bursts \citep{weinberg_2006_aa,guichandut_2023_aa}, 
Type Ia supernovae \citep{shen_2018_aa,collins_2022_aa}, and 
He-rich subdwarf O stars \citep{werner_2022_aa,miller-bertolami_2022_aa}. 
Helium burning also impacts several classes of distribution functions,  
such as the black hole mass distribution function \citep{fryer_2001_aa,sukhbold_2018_aa,sajadian_2023_aa} 
including any mass gaps based on the pair-instability mechanism in the evolution of 
massive stars \citep{fowler_1964_aa, woosley_2002_aa, farmer_2019_aa, farmer_2020_aa, renzo_2020_aa, marchant_2020_aa, farag_2022_aa}.

He burning is triggered by the 3$\alpha$ process releasing 7.5\,MeV in fusion energy and producing $^{12}$C \citep{hoyle_1954_aa,eriksen_2020_aa,kibedi_2020_aa,cook_2021_aa}. 
This is a unique process, setting stringent conditions for helium ignition.
The 3$\alpha$ process is followed by the $\alpha$ capture reaction $^{12}$C($\alpha$, $\gamma$)$^{16}$O,
converting the $^{12}$C into $^{16}$O \citep{deboer_2017_aa,mehta_2022_aa,shen_2023_aa}. 
These two isotopes are the principal products of He burning.
In addition, nearly all of a star's initial CNO abundances in the stellar interior are converted to $^{22}$Ne at the onset of He burning \citep{timmes_2003_aa, howell_2009_aa, bravo_2010_aa, blondin_2022_aa, meng_2023_aa}.
This marks the first time in a star's life where the core becomes neutron rich. We follow the convention that $^{22}$Ne is the ``metallicity'' of a carbon-oxygen (CO) white dwarf (WD).

The interiors of CO WDs are, in principle, the best probe of the ashes of He burning.
A goal of WD seismology is to characterize the chemical profiles of principal products of He burning
\citep{metcalfe_2001_aa,  metcalfe_2002_aa, fontaine_2002_aa, straniero_2003_aa,  metcalfe_2003_aa, giammichele_2017_ab, 
de-geronimo_2017_aa, giammichele_2018_aa, de-geronimo_2019_aa, corsico_2019_aa, pepper_2022_aa, giammichele_2022_aa, corsico_2022_aa, romero_2023_aa}
and the chemical profile of the trace $^{22}$Ne metallicity \citep{camisassa_2016_aa, giammichele_2018_aa, chidester_2021_aa, althaus_2022_aa}.

Furthermore, regions within a CO WD model that burn helium radiatively during its prior evolution can offer potential constraints on the He burning nuclear reaction rates.
For example, \citet[][hereafter C22]{chidester_2022ApJ...935...21C} found that certain trapped adiabatic g-modes in WD models 
may provide a pulsation signature that constrains the experimental \COrate\ reaction rate probability distribution function. 
These signature g-modes were shown to resonate 
with the region of the CO WD model that underwent radiative He burning during its previous evolution. The innermost boundary of this resonant cavity 
corresponds to the molecular weight gradient at O$\rightarrow$C chemical transition, and the outermost boundary to the molecular weight C$\rightarrow$He chemical transition. 
The resonating region encompasses $\simeq$\,0.2\,\Msun\ of a typical $\simeq$\,0.6\,\Msun\ WD model.
C22 cautioned that the chemical structure and resulting pulsation spectrum 
is sensitive to 
the width of the O$\rightarrow$C transition \citep{corsico_2002_aa, salaris_2017_aa, pepper_2022_aa}, 
the experimental 3$\alpha$ reaction rate probability distribution functions \citep{deboer_2017_aa, kibedi_2020_aa, schatz_2022_aa},
convective boundary mixing processes during core He depletion \citep{salaris_2017_aa, anders_2022_aa}, and 
the number of thermal pulses during the Asymptotic Giant Branch (AGB) phase of evolution \citep{de-geronimo_2017_aa, pepper_2022_aa}.

Modeling convective boundary mixing processes at the convective-radiative interface during core He burning in low- and intermediate-mass stellar models is currently uncertain 
\citep{herwig2002, salaris_2017_aa,jermyn_2022_aa,anders_2022_aa, blouin_2023_aa}.
Convective overshoot occurs because the convective boundary is not the location where convective velocities are zero, 
but the location where the buoyant acceleration of the fluid is zero.
An order--of-magnitude expression $\Delta x = u \Delta t$ provides an estimate for how far convective motions overshoot \citep{anders_2022_aa}.
Here $\Delta x$ is the overshoot distance, $u$ is the convective velocity, and 
$\Delta t \simeq 1/N$ where $N$ is the \BV\ frequency
in the stable region. There is disagreement on how to calculate $\Delta x$, but this estimate 
broadly shows $\Delta x \ll H_P$ in stellar environments, where $H_P$ is the pressure scale height.
The exponential overshoot parameterization \citep[e.g., ][]{herwig2002} is frequently implemented in 1D models to describe this convective boundary mixing process, treating $\Delta x$ as a free parameter. 
The values of $\Delta x$
needed to match the gravity modes found in Slowly Pulsating B-type stars \citep{pedersen_2021_aa} suggest $\Delta x / H_P \simeq 0.1$, which is larger than 3D hydrodynamical simulations of low Mach number flows at stable interfaces indicate \citep{korre_2019_aa, blouin_2023_aa}.

The injection of fresh He into the convective core enhances the rate of energy production by the $^{12}$C$(\alpha,\gamma)^{16}$O reaction rate, increases the central $\oxygen$ mass fraction \citet[e.g., ][]{de-geronimo_2017_aa}, and modifies the lifetime through this phase of evolution.
The resulting increase in the radiative gradient can also lead to rapid growth in the convective He core boundary (a ``breathing pulse'').
A consensus on breathing pulses being physical or numerical has not yet been reached \citep{caputo_1989_aa, cassisi_2003_aa, farmer_2016_aa, constantino_2017_aa, paxton_2019_aa}. 

C22 found a pulsation signature of the  \COrate\ reaction rate probability distribution function using evolutionary models that purposely excluded overshooting. 
This article is novel in analyzing whether or not pulsation signals of the \COrate\ reaction rate probability distribution function
still exist when overshooting at the inner convective-radiative interface during core He burning (CHeB) is included in the models' evolution history. Here, the inner convective-radiative interface is the transition from the convective core to the exterior radiative layer.
Section~\ref{sec:models} describes our models, 
\S\ref{sec:results} analyzes our models,
\S\ref{sec:discussion} discusses our results,
and we summarize our findings in \S\ref{sec:summary}.
Appendix A lists the microphysics used, and 
Appendix B discusses variations with the number of isotopes in the reaction network and with the temporal resolution of our models.

\section{Stellar Evolutionary Models}
\label{sec:models}

We define the term ``model'' to mean an evolutionary sequence that begins at the pre-main sequence, progresses through CHeB, and terminates as a cold WD. We define the term ``snapshot'' to mean a specific instance in time or phase of evolution within a model, and the term ``set'' to mean a suite of models or snapshots that have identical input physics except for the value of the \COrate\ reaction rate.

We use \MESA\ version r15140 
\citep{paxton_2011_aa,paxton_2013_aa,paxton_2015_aa,paxton_2018_aa,paxton_2019_aa,jermyn_2023_aa} to build 2.1\,\Msun, 
$Z$\,=\,0.0151 metallicity, $Y$\,=\,0.266 He mass fraction, nonrotating models at the pre-main sequence. 
We adopt the AGSS09 \citep{asplund2009ARA&A..47..481A} abundances and use a 23 isotope nuclear reaction network with $^{22}$Ne being the heaviest isotope\footnote{A comparison to a 30 isotope network is given in Appendix B.}.  
Our models employ \MESA's \code{Henyey} mixing-length theory (MLT) of convection option, with an MLT parameter of $\alpha$\,=\,1.5. \revision{This is consistent with the value used in C22.}
We use the Ledoux criterion, and the predictive mixing scheme. 
Additional details of the $\MESA$ microphysics are listed in Appendix A.

\begin{figure*}[!ht]
  \includegraphics[width=\textwidth]{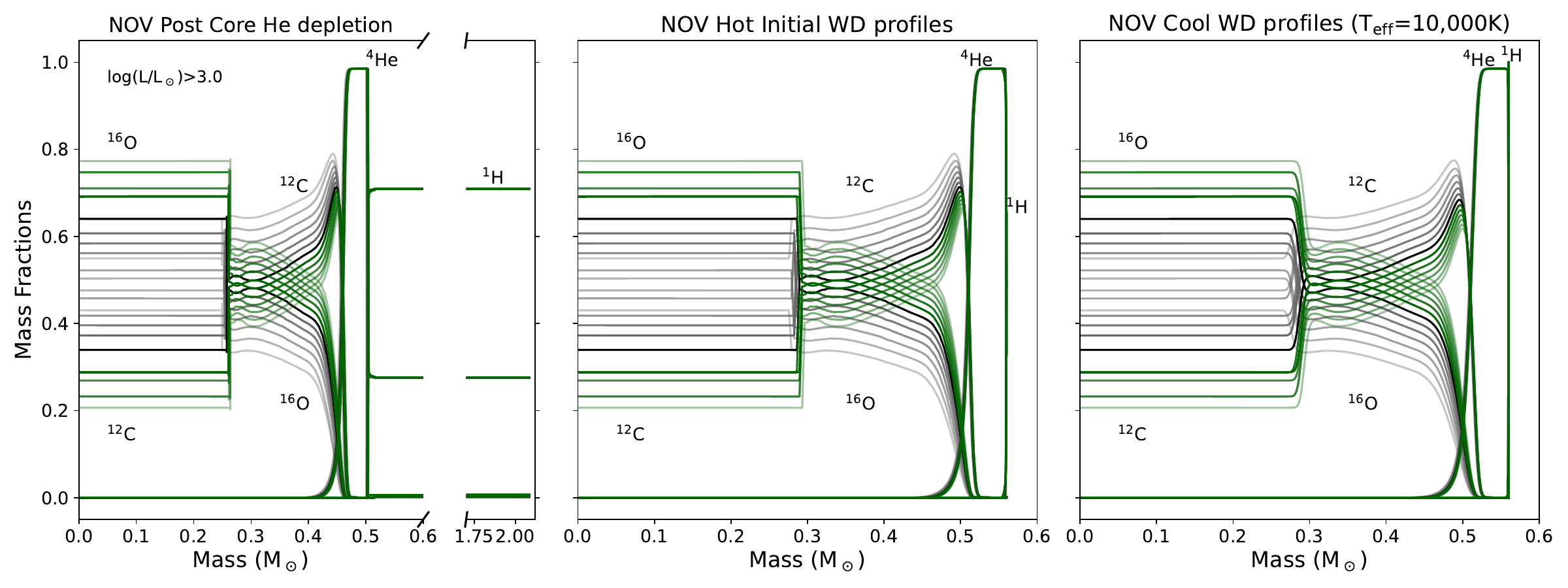}
  \includegraphics[trim={3.5cm 0.5cm 4cm 0cm},clip,width=\textwidth]{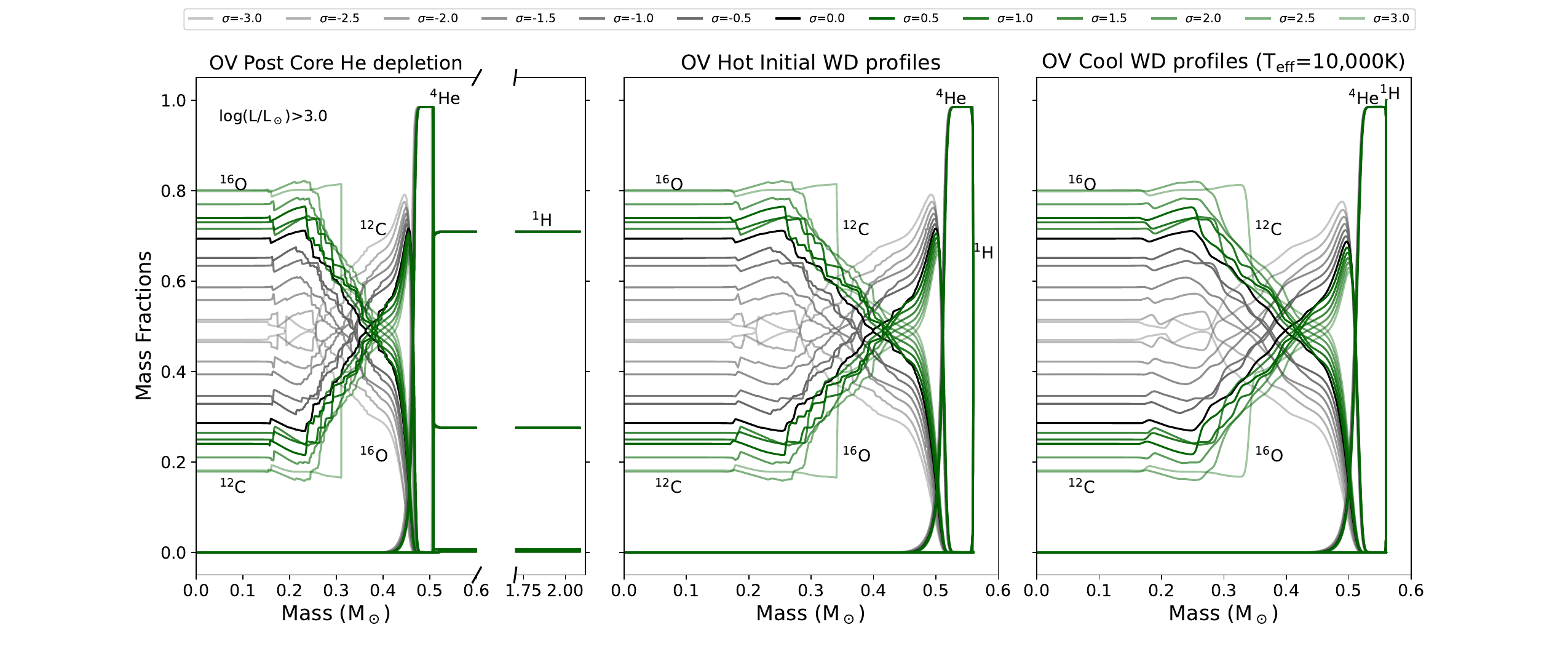}
  \caption{\textbf{Top:} Mass fraction profiles without overshooting (NOV) during the CHeB phase.  \textit{Left:} Mass fraction profiles after core He-depletion, terminated prior to the first thermal pulse at $\log(L/L_\odot)>3.0$. \textit{Middle:} Mass Fraction profiles at the first \code{wd\_builder} model step. These profiles have been shaved of their excess H envelope prior to running on the WD cooling track.  These are the initial hot WD profiles.  \textit{Right:} Mass fraction profiles when the models have cooled to $\Teff$ = 10,000 K.  The smoothness in the profiles reflects the element diffusion processes included in the calculation. \textbf{Bottom:} Mass fraction profiles with core overshooting (OV) during the CHeB phase, in the same format as above.  Green curves represent positive $\sigma_i$ \COrate\ reaction rates, grey curves represent negative $\sigma_i$ \COrate\ reaction rates.  For both positive and negative $\sigma_i$, the shading grows fainter the further $\sigma$ is from the standard rate ($\sigma=0$; black curve).}
  \label{fig:nov_ov_profiles}
\end{figure*}

As in C22, we span the current experimental $\COrate$ reaction rate probability distribution function \citep{deboer_2017_aa, mehta_2022_aa} from $\sigma=-3.0$ to $\sigma=+3.0$ in 0.5$\sigma$ steps, totaling to 13 $\sigma_i$ reaction rates; each  model is prescribed one such $\sigma_i$ $\COrate$ reaction rate value for its evolution.
We calculate one set of models without overshooting (NOV), and a second set with overshooting  (OV) at the inner radiative-convective interface during the CHeB phase.
Hence, each evolutionary model differs only in its $\sigma_i$ $\COrate$ reaction rate, and NOV or OV mixing prescription.  This yields 26 individual stellar evolutionary models; 13 for the NOV set and 13 for the OV set.    For $i=(-3.0, -2.5,...,+2.5, +3.0)$, we use $\sigma_i$ and $\sigma=i$ interchangeably to reference a given $\sigma$ from the \COrate\ reaction rate probability distribution function.  

After CHeB, the models evolve until $\log(L/L_\odot)=3.0$, prior to the first thermal pulse on the AGB.  At this snapshot, we interrupt the evolution of each model. All models at this snapshot thus have a C$\rightarrow$He transition at nearly the same mass location.  We use this snapshot to construct H-dominated atmosphere (DA) WDs by removing the H envelopes until $\log$(M$_H$/M$_*$)$<-$3.5.   
 The resulting composition profile structures are used to build 0.56\,$\Msun$ ab-initio WD models with \code{wd\_builder}, as done in C22.  These WD models evolve until $\Teff$\,=\,10,000~K.  We discuss the reasoning for constructing the WDs from the post-CHeB $\log(L/L_\odot)=3.0$ snapshot in the following section.

We utilized version 6.0.1 of the $\GYRE$ code \citep{townsend_2013_aa,townsend_2018_aa} to compute the adiabatic pulsations of our WD models throughout their respective cooling tracks (from $\sim 50,000$~K to $10,000$~K).  \revision{We tracked the pulsations for the entire WD cooling track to observe the evolution of the adiabatic modes.  Further, this was the most convenient way to auto-implement pulsation calculations for multiple models (i.e. we did not have to post-process the pulsation calculations over a specifed $\Teff$ range for each of the 26 models).  We emphasize that the computed pulsations are adiabatic, and that the observed instability strip for DAV WDs spans only from $\sim 13,000$ K to $\sim 10,000$ K.} The inlist parameters were set to search for modes of harmonic degrees $\ell=1,2$ and radial orders $n\leq25$, where our models were assumed to be non-rotating, hence only $m=0$ azimuthal orders were present. For the adiabatic mode analysis, we employed the fourth-order Gauss-Legendre collocation difference equation scheme \citep{gauss-legendre_iserles1996first,townsend_2013_aa,townsend_2018_aa}.


Details of the $\MESA$ models and $\GYRE$ oscillation parameters are in the files to reproduce our results at
at doi:\dataset[10.5281/zenodo.8126450]{[https://doi.org/10.5281/zenodo.8126450}.

\subsection{Core Overshooting prescription during the CHeB}

During the CHeB phase, we use the following core overshooting parameters in the \MESA\ inlist for the OV set:

\begin{itemize}\tightitems
\item[] 
\texttt{! overshoot}
\item[]
\texttt{min\_overshoot\_q} = 1d-3  
\item[]
\texttt{overshoot\_scheme(1)} = `exponential' 
\item[]
\texttt{overshoot\_zone\_type(1)} = `any' 
\item[]
\texttt{overshoot\_zone\_loc(1)} = `core' 
\item[]
\texttt{overshoot\_f(1)} = 0.016 
\item[]
\texttt{overshoot\_f0(1)} = 0.008 
\item[]
\texttt{overshoot\_mass\_full\_on(1)} = 0.01 
\item[]
\texttt{overshoot\_mass\_full\_off(1)} = 0.4 
\end{itemize}

Details of the specific parameters are described in the \MESA\ documentation\footnote{\url{https://docs.mesastar.org/en/latest/}}.  
We choose the conventional \cite{herwig2002} value of \verb|overshoot_f(1)=0.016|.   
This parameter sets the fractional distance of $H_p$ to overshoot at the $\nabla_{ad}=\nabla_{rad}$ interface, for the order of magnitude estimate given in the introduction, $\Delta x = f_0\cdot H_p$. 


The trapped mode seismic signatures found in C22 were resonating most with the region that underwent radiative He burning, defined as R2. Their inner boundary of R2 is near the molecular weight gradient at the 
O$\rightarrow$C transition (the ``O drop") and their outer boundary is near the C$\rightarrow$He transition. Mode trapping is sensitive to the location of both of these boundaries because they define the width of the resonant cavity. 

One approach to analyzing the sensitivity 
of the R2 trapped mode signatures is to fix one boundary and vary the other boundary. We fix the R2 outer boundary by excluding variations imposed from the thermal pulse history, hence the interruption at the post-CHeB $\log(L/L_\odot)=3.0$ snapshot for all models.  The phenomena that happens during the AGB phase is another source of model uncertainty.  \cite{alfred2023arXiv230311374G} found that early post-AGB pulsations can cause rapid growth of an instability that drives a super-wind which can shed much of the outer layers in a few years.  Further, their 2.0 $\Msun$, Z=0.02 model shows a dynamic evolutionary track, especially during the AGB, that is similar to the models in this article. \cite{alfred2023arXiv230311374G} summarizes that while the preliminary results show promise on future AGB and post-AGB phenomenon, there are currently more questions than answers.  We therefore leave the thermal pulse history and the particular envelope ejection phenomena on the AGB to future studies, and freeze the outermost R2 boundary before the first thermal pulse occurs.  In this vein, we isolate the sensitivity of the R2 region to its inner boundary, and specifically address how core overshooting influences the pulsation signatures for the $\COrate$ reaction rate probability distribution function.

We end this section by stating we are not advocating for a specific evolutionary model or overshooting scheme. 
Rather, we are exploring one approach to quantifying the coupled uncertainty between the \COrate\ reaction rate probability distribution function and a common overshooting model.

\section{Results}
\label{sec:results}
\subsection{Evolution of Composition Profiles}

\begin{figure}[ht!]
    \centering
     \includegraphics[trim={0.1cm 0cm 0.1cm 0cm},clip,width=0.48\textwidth]{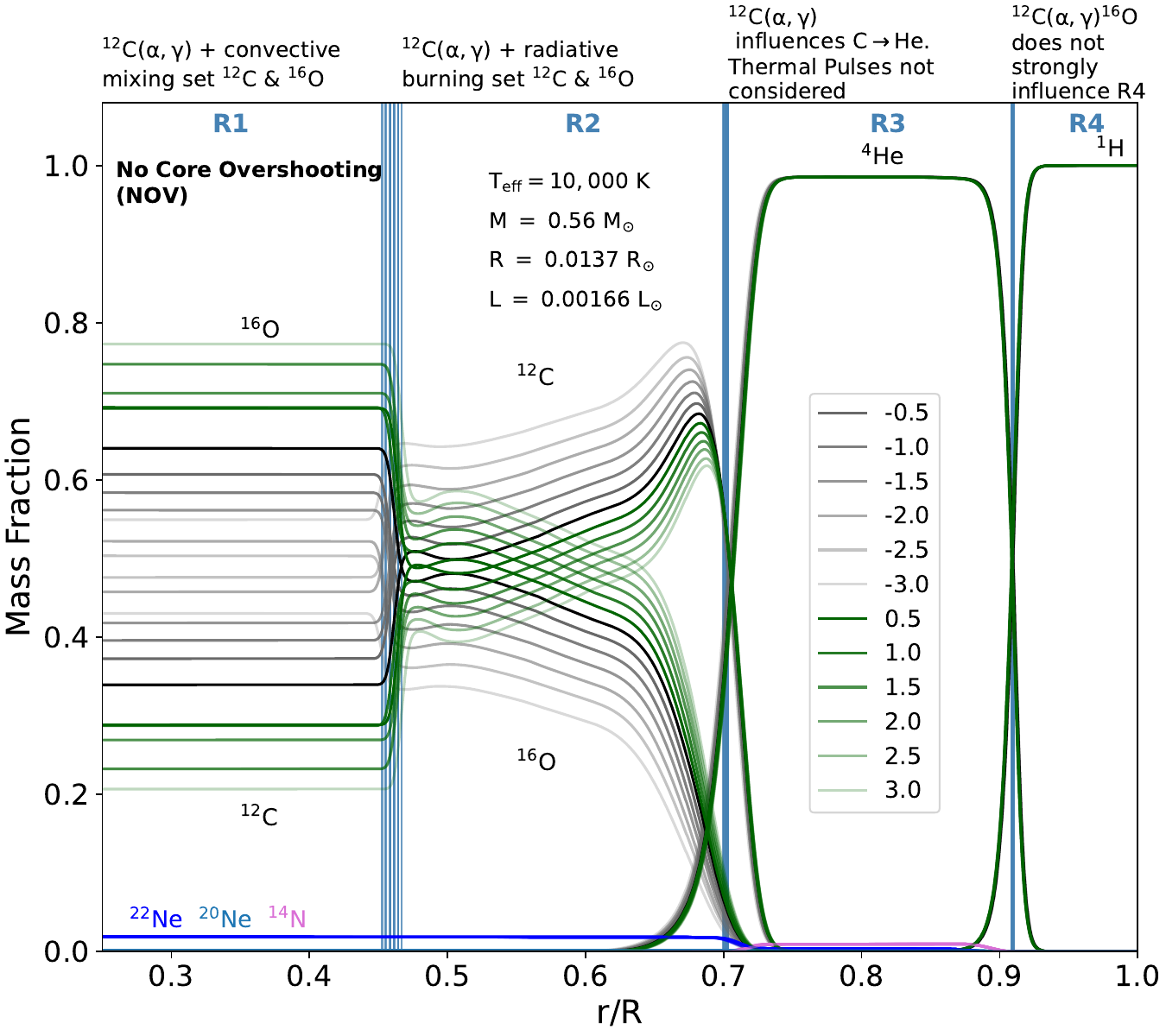}    
     \includegraphics[trim={0.1cm 0cm 0.1cm 0cm},clip,width=0.48\textwidth]{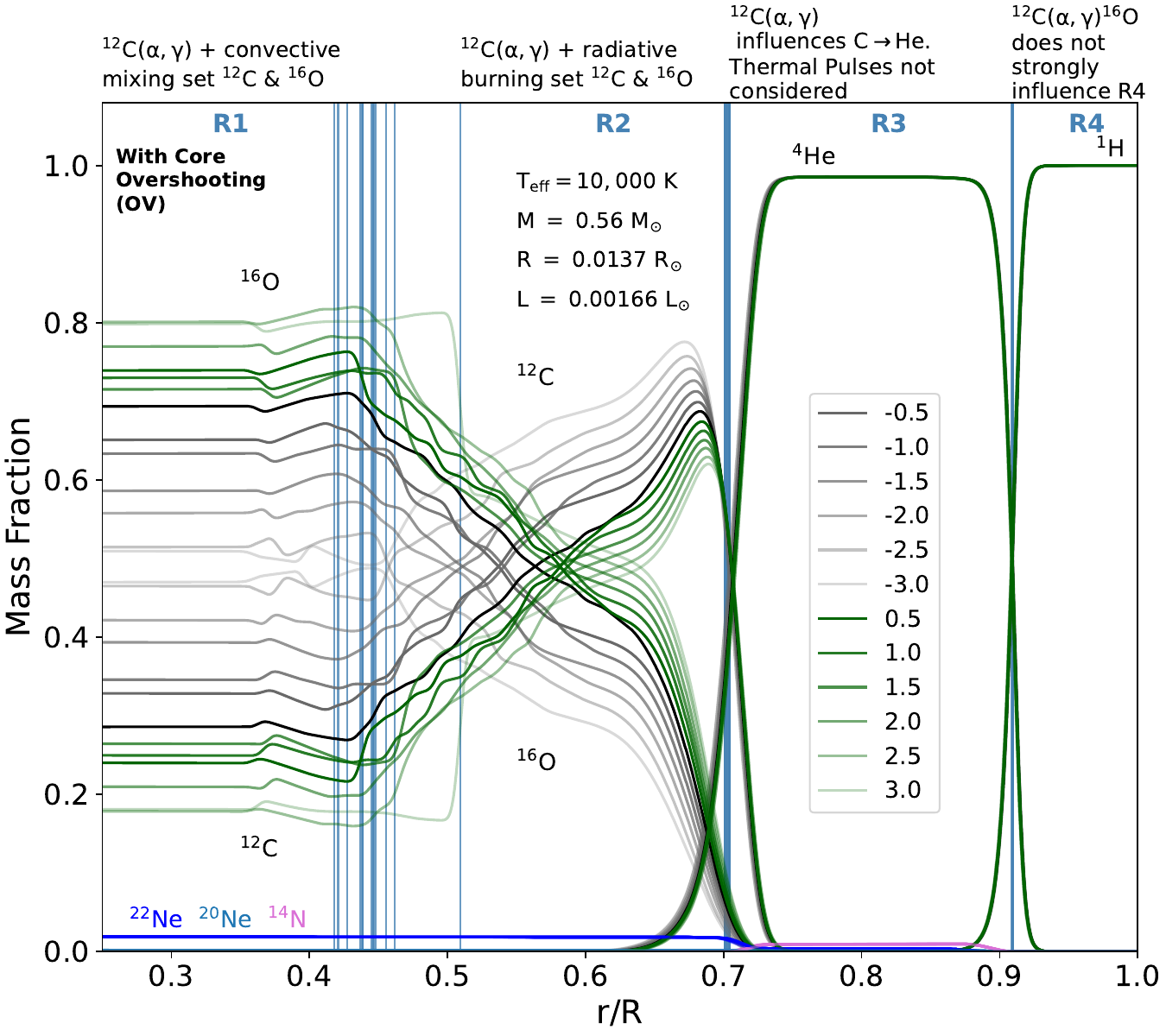}
    \caption{
    Composition profiles after the NOV (top) and OV (bottom) models cool to $\Teff$ = 10,000 K. Region boundaries are indicated by vertical blue lines, and the $\sigma_i$ colors are the same as in Figure~\ref{fig:nov_ov_profiles}.}
    \label{fig:wd_profiles}
\end{figure}

\begin{figure*}[ht!]
\begin{interactive}{js}{figures/interactive_kipp_zip_folder.zip}
\begin{center}
\includegraphics[width=0.48\textwidth]{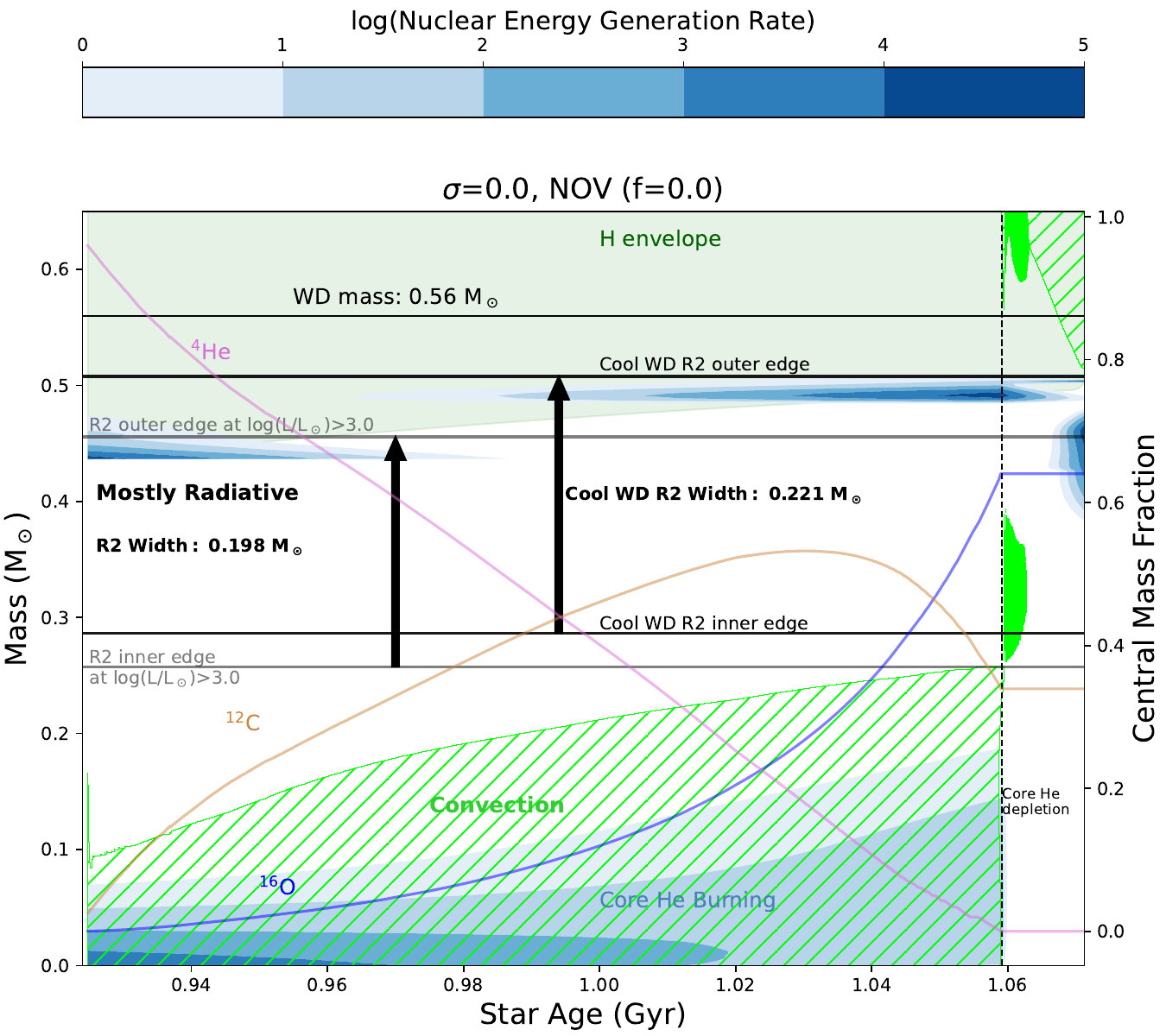}
\includegraphics[width=0.48\textwidth]{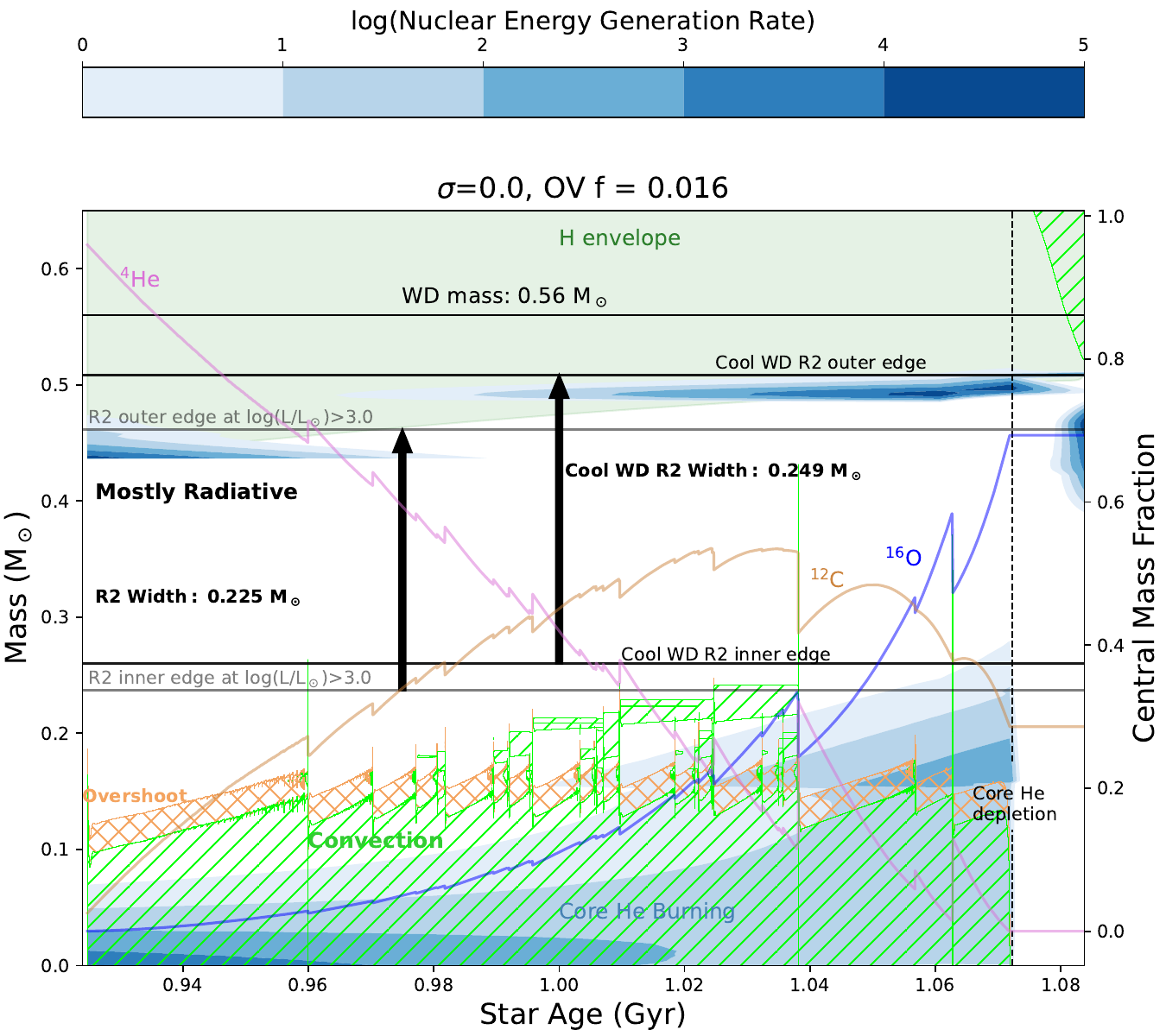}

\end{center}
\end{interactive}

\caption{Kippenhahn diagrams for the NOV (left) and OV (right) $\sigma=0.0$ models.  
The x-axis is the respective stellar model's age, the left y-axis is the mass coordinate, the right y-axis is the central mass fraction of the isotopes. 
Bright green areas represent convection, blue shaded regions depict nuclear burning (see colorbar), white areas represent radiation, and yellow-gold areas represent overshooting (right figure). The light green area shows the hydrogen envelope.  The solid pink curve is the central $\helium$ mass fraction, the solid dark blue curve is the central $\oxygen$ mass fraction, and the dark yellow curve is the central $\carbon$ mass fraction.  The dashed line shows core helium depletion.  The evolution was terminated when $\log(L/\Lsun)>3.0$ for all stellar models, and the figures are plotted until that point. Annotated is the radiative R2 region's edges and widths.  An interactive figure is provided in the online version. Its functionality compares the NOV and OV Kippenhahn diagrams for any given $\sigma_i$ $\COrate$ reaction rate.  To use the interactive figure, click on the `sigma' slider at the top, and slide through the indices 0 through 12. Sliding through each index provides the user the side by side comparison of the Kippenhahn diagrams for the respective $\sigma_i$ $\COrate$ reaction rate used in the evolution. Index 0 starts at $\sigma_{-3.0}$ and index 12 ends at $\sigma_{3.0}$.  }
\label{fig:kipps}
\end{figure*}

Figure~\ref{fig:nov_ov_profiles} shows the mass fraction profiles for both sets at three evolutionary snapshots.  The top row shows the mass fraction profiles for the NOV set and the bottom row shows the mass fraction profiles for the OV set.  The left most column  
shows the mass fraction profiles at the post-CHeB $\log(L/L_\odot)>3.0$ snapshot. At this point, our models have not lost much mass and are all $\sim2.1 \ \Msun$.  The middle column shows the mass fraction profiles after removing the H envelopes until $\log$(M$_H$/M$_*$)$<-$3.5. This snapshot shows the initial hot WD profiles, after completing one model step in \code{wd\_builder}.  The profiles shift slightly in mass location, but the overall composition structure only differs from the left panel in the thickness of the H envelope.  The right column is the final snapshot of the mass fraction profiles, when the models reach $\Teff=10,000$ K.  Diffusion was included on the WD cooling track and leads to the smoothness of the profiles in this column.

Figure ~\ref{fig:wd_profiles} accentuates the differences between the NOV (top) and OV (bottom) mass fraction profiles for the final WD structures (right column of Figure~\ref{fig:nov_ov_profiles}).  Here, we show the abundance in mass fraction with respect to fractional radius $r/R$. We partition the WDs' composition profiles into four regions: R1, R2, R3, and R4.  This is similar to that done in C22.  The regions are defined to estimate trapping (resonant) zones.  Boundaries for mode trapping are typically near composition transitions because they generally have large mean molecular weight gradients.  This may lead to partial reflections for a resonant mode(s), ``trapping" it within the local cavity \citep{brassard_1991_aa,winget_1981_aa}.  The Ledoux B profile (henceforth $B$) captures composition gradients and can estimate trapping regions.  We use $B$ as our primary guide to define the region boundaries for a given model. The R1-R2 boundary is set at the first local maximum in $B$ that occurs after reaching peak $\oxygen$ in a given model's chemical profile.  The R2-R3 boundary is set at the second local maximum in $B$.  The R3-R4 boundary is set at the location where $\rm X(^1H)>X(^4He)$. 

In both NOV and OV sets, $\sigma_i$ impacts the magnitude of the $\rm ^{16}$O and $\rm ^{12}$C profiles in R1. Core overshooting changes the structure of these profiles, especially at $r/R \sim 0.37$ where the flatness of the profiles becomes disrupted. This is due to additional He fuel ingested during CHeB, from overshooting and/or convection. The fuel ingestion from overshooting and convection is a coupled effect and specific to each $\sigma_i$ model.  After $r/R \sim 0.37$, there is some overlap in the profiles that perturbs the proportional trend with $\sigma_i$. 

For both sets, the first group of vertical blue lines marks the R1-R2 boundary, with each line representing a given $\sigma_i$. The NOV set shows a steep composition gradient at the R1-R2 boundary, and the R1-R2 location is nearly the same for all $\sigma_i$. There is greater variance in the R1-R2 location for the OV set.  Further, core overshooting has softened the $\oxygen$ and $\carbon$ gradients, and the disruption of the profiles' regularity with $\sigma_i$ continues into the start of the R2 region.  At $r/R\sim0.6$, the proportionality of $\sigma_i$ to the $\carbon$ and $\oxygen$ profiles is restored.

By design from stopping at the first thermal pulse, the R3 and R4 regions are almost identical between the NOV and OV sets.  These regions are least affected from mixing processes in the core (e.g. overshooting).

In Figures \ref{fig:nov_ov_profiles} and \ref{fig:wd_profiles}, the OV chemical profiles show a non-constant structure from overshooting during CHeB in the O dominated central core (below $\simeq$0.4\,\Msun). While element diffusion is included during the white dwarf cooling phase, these chemical profiles may be further flattened by mixing processes not considered in this study such as time-dependent convection \citep{jermyn_2023_aa}, rotationally induced mixing, semiconvection, thermohaline mixing, \revision{or first-order phase separation of the CO mixture \citep{bauer_2023_aa}.}

 \subsection{Evolutionary differences after the main-sequence}
How do the final WD profiles for the NOV and OV sets in Figure~\ref{fig:wd_profiles} relate to their respective CHeB evolution histories?  Figure~\ref{fig:kipps} shows the Kippenhahn diagrams for the $\sigma$\,=\,0.0 models for NOV (left) and OV (right).  This figure shows the CHeB phase until the $log(L/\Lsun)>3.0$ termination point, spanning $\simeq$\,0.93--1.10~Gyr.  During this period the total mass of our models is $\simeq$\,2.1~\Msun, but we show only the innermost  $\simeq$\,0.65~\Msun to capture the evolution history that ultimately defines the CO WDs.

There are immediate differences between the NOV and OV CHeB evolution histories for the $\sigma=0.0$ models.  These differences are similar for any given $\sigma_i$ models, and a link to an interactive figure is provided in the online journal to see each rate's OV vs. NOV comparison in greater detail.  

For the NOV set, we see gradual growth of the convective core throughout the CHeB phase; the noted central mass fraction isotopes smoothly deplete/grow to reach their final mass fractions; the convective cores have no apparent splitting during the CHeB phase.  Further, there is a pure radiative zone throughout the CHeB history.  In comparison, the OV set shows convective cores that ebb and flow in their extent, in a saw-tooth like manner; overshooting extends past the inner convective core in a fairly consistent mass length; the OV central mass fraction isotopes ebb and flow symmetrically with the mixing phenomena at any given time.  

We also see splittings of the convective core in the OV set.  These splittings were not observed in any of the NOV models during the CHeB phase. We presume they are a result of overshoot inclusion. This introduces ``pollution" to the pureness of the radiative burning zone, which becomes the R2 region of the WD.  The pollution is seen by observing that some of the split-convection zone surpasses the $\log(L/\Lsun)>3.0$ R2 inner edge boundary.  This boundary becomes the inner edge of R2 in the cool WDs.  The amount of convective pollution within the OV set is minor for $\sigma_{0.0}$, but varies with $\sigma_i$.  
Figure~\ref{fig:kipps} qualifies R2 as ``Mostly Radiative" for the NOV set due to localized, short-lived, subtle convective occurrences between $\simeq$\,0.30--0.35 \ $\Msun$ near core He depletion energetics.  Composition profiles are less sensitive to mixing after CHeB is complete. Any convective pollution from these brief convective periods in the NOV set are insignificant compared to the convective pollution introduced in the OV set. 

For both sets, nuclear burning primarily takes place within the convective core. Both sets also show similar burning regions in the mantle outside the core, in the radiative zone.  Near the end of core He depletion, nuclear burning in the core extends past the convective and overshooting core regions in the OV set, and burns into the radiative zone.  This is not seen in the NOV set.

\subsection{WD Adiabatic Pulsation Analysis}
How do these evolutionary and WD structural differences impact the WD \COrate\ reaction rate pulsation signatures?  We first stress the importance of the NOV models' R2 pure radiative zone during the CHeB. The trapped mode $\sigma_i$ signature found in C22  resonates the most with this region.
We want to determine if this signature, or any other $\sigma_i$ pulsation signature, exists when overshooting is considered at the inner R2 boundary during CHeB.  First we compare the NOV WD pulsation signatures in this work to those in C22.  

\begin{figure}[!htb]
     \includegraphics[width=0.5\textwidth]{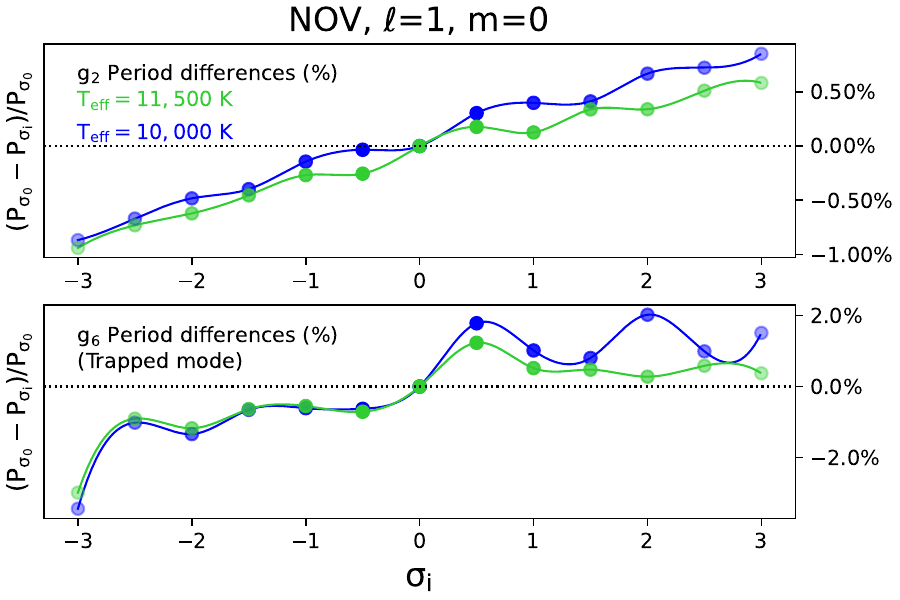}
      \includegraphics[width=0.23\textwidth]{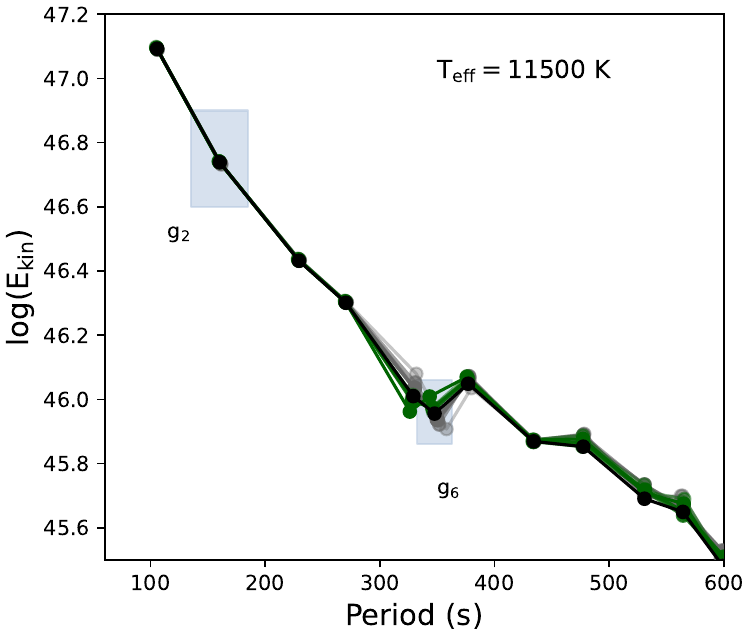}
    \includegraphics[width=0.23\textwidth]{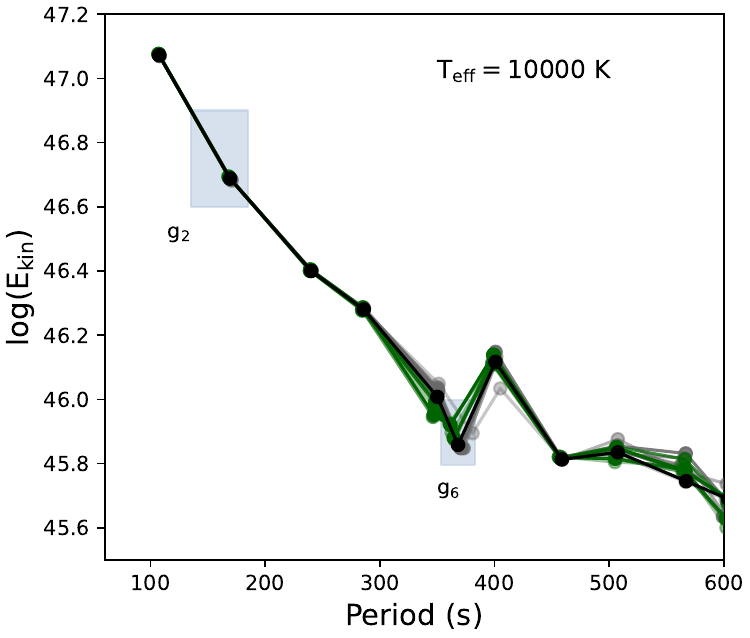}
    
    \caption{\textit{Top two:} NOV set's signature adiabatic pulsation modes shown at $\Teff = 11,500$ (bright green), and $\Teff=10,000$ (blue) K respectively.  The first panel shows the signal from the $g_2$ mode; the second shows the signal from the $g_6$ trapped mode.  \textit{Bottom:} The kinetic energy diagrams for all $\sigma_i$ at $\Teff=11,500$ K (left) and $\Teff=10,000$ K (right) respectively. The green dots/lines represent $E_{kin}$ for $\sigma_i>0$; grey for $\sigma_i<0$; black for $\sigma=0$.  The shading of color gets fainter the further away from $\sigma=0$. }
    \label{fig:nov_signals}

\end{figure}

\begin{figure}[!htb]
    \centering
    \includegraphics[width=0.48\textwidth]{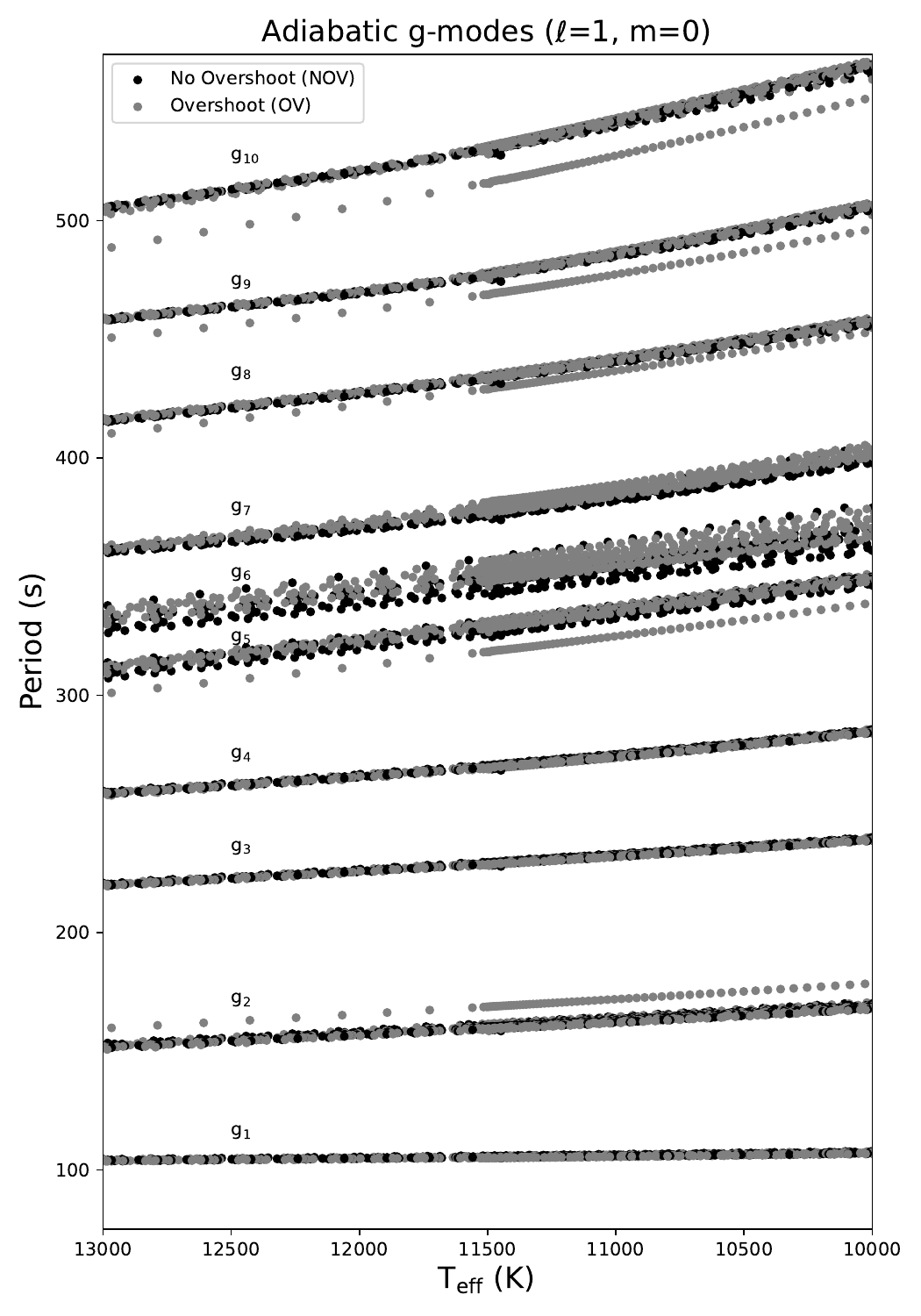}
    \caption{Pulsation periods as function of $\Teff$ for the NOV (black) and OV (grey) model sets.}
    \label{fig:raw_periods}
\end{figure}

\subsection{NOV set vs. C22}
\revision{In this section we briefly describe the main differences between the NOV and C22 models.}  The models in C22 used a 30 isotope chemical network compared to the 23 isotope network used here. See Appendix B for a comparison. Also, the temporal resolution was greater in C22, especially through CHeB.  The most important difference in the NOV models is that we terminated the evolution prior to the first thermal pulse; the models in C22 continued the evolution through the thermal pulse phase of evolution. The overall composition structure of the R1 and R2 regions in our NOV models are quite similar to those in C22.

The NOV set of models in this work found two WD g-mode signals for $\sigma_i$ rather than one.  This is shown in the top two panels of Figure~\ref{fig:nov_signals}.  Both panels show snapshots of the percent period differences as a function of $\sigma_i$, at $\Teff=11,500$ K (bright green) and $\Teff=10,000$  K (blue) respectively.  The y-axis label defines the period differences as $(P_{\sigma_0}-P_{\sigma_i})/P_{\sigma_0}$. That is, they are normalized to the  pulsation periods of the $\sigma=0$ NOV model. The first panel is the signal from $g_2$ and the second is the signal from $g_6$.  In C22, 
the g-mode signature was a trapped mode.  Trapped modes are identified from local minima in the kinetic energy diagram \citep{winget_1981_aa, brassard_1991_aa}.  The NOV kinetic energy diagrams for all $\sigma_i$ at these snapshots are shown in the bottom left and right panels of Figure~\ref{fig:nov_signals}, following Equation 2 in C22 
\citep{unno_1989_aa,corsico_2002_aa}.  The figure caption explains the coloring for $\sigma_i$. At $\Teff=11,500$~K (bottom left panel), the first apparent trapped mode occurs at $g_6$ for all $\sigma_i$, with the exception of $\sigma=0.5$, which has its first local minimum of $E_{kin}$ at $g_5$.  By $\Teff=10,000$ K (bottom right panel), all $\sigma_i$ have the first local minimum in $E_{kin}$ at $g_6$, including $\sigma=0.5$.  This is important as $g_6$ is one of our signature modes for $\sigma_i$.  These findings are in overall agreement with C22. 
The trapped $g_6$ mode signature is not  linear with $\sigma_i$, but overall shows $\sigma_i<0$ to have longer periods than $\sigma=0.0$, and $\sigma_i>0$ to have shorter periods than $\sigma=0.0$.  
The R2 contribution to the $g_6$ period in our NOV models was $\sim 25$\%. Other regions equally contributed between $\sim 20-30$\%, meaning that the trapped mode from our NOV set is more equitably trapped among the four regions.  Thus, its credibility from R2 isn't as strong as in C22. 
Nonetheless, it is not a negligible contribution and can still serve as a viable probe for $\sigma_i$. 

Our other g-mode signal, $g_2$, does not appear to be trapped by definition (see other highlighted mode in bottom of Figure~\ref{fig:nov_signals}).   However, the $g_2$ period differences are directly proportional to $\sigma_i$ (first panel of Figure~\ref{fig:nov_signals}).  This suggests that $g_2$ is likely distinguishing CO features in the inner regions better than other g-modes. The additional $g_2$ signal
was either recovered or contrived as a consequence of excluding the thermal pulse history in the evolution. This was the only procedural difference between our models and those in C22. 
The direct impact of this procedural difference is expressed by the nearly uniform \carbon\ and \helium\ profiles after the C$\rightarrow$He transition (see Figure~\ref{fig:nov_ov_profiles}).  
C22 showed variations in these profiles that stemmed from variations in the thermal pulse histories.  Eliminating such chemical variations near the R2-R3 interface can placate the g-modes' sensitivity to the R3 and R4 regions, especially for low-order g-modes such as $g_2$.  Figure 9 in  
C22 shows $g_2$ distinguishes $\sigma_i$ in their thinner atmosphere sequence of models.  Thinner atmospheres may also lessen sensitivities to outer regions, allowing lower-order g-modes like $g_2$ to probe deeper into the CO interior.   We therefore suspect $g_2$ is a viable probe for $\sigma_i$ if there are uniform composition profiles at the R2-R3 boundary, and/or thinner WD atmosphere models.

We conclude that our NOV pulsation signature results are overall consistent with  C22; 
we find certain low-order adiabatic WD g-modes which probe the \COrate\ reaction rate probability distribution function.  With our two signature modes established, we now discuss the impact that overshoot inclusion has on these pulsation signatures.

\subsection{Detailed Analysis of Differences}
We first show the pulsation periods as a function of surface temperature for all $\sigma_i$ models in Figure~\ref{fig:raw_periods}.  Black dots mark the NOV periods and grey dots mark the OV periods. G-modes with radial orders $n=1-10$ are annotated, all for $\ell$\,=\,1.  Figure~\ref{fig:raw_periods} shows that there are differences in the periods between the NOV and OV sets, but there is no global systematic offset; the differences between the OV and NOV periods for any given g-mode is random.  This is the case even when $\sigma_i$ is constant.  We find that $g_6$ shows the largest spread in the periods of the models. Further, the kinetic energy diagrams for all models show that $g_6$ was a trapped mode by $\Teff=10,000$ K for every model, regardless of the $\sigma_i$, NOV/OV prescription.  Since $g_6$ is one of the signals for $\sigma_i$ in the NOV models, we point out this feature in Figure~\ref{fig:raw_periods}. We will touch on the cause of the larger spread later, but now focus our attention on the detailed pulsation properties of the signature $g_2$ and $g_6$ modes.

\begin{figure*}[!htb]
\begin{interactive}{js}{figures/4panel_interactive.zip}
\begin{center}
\includegraphics[trim={0cm 1cm 1.5cm 1cm},clip,width=0.44\textwidth]{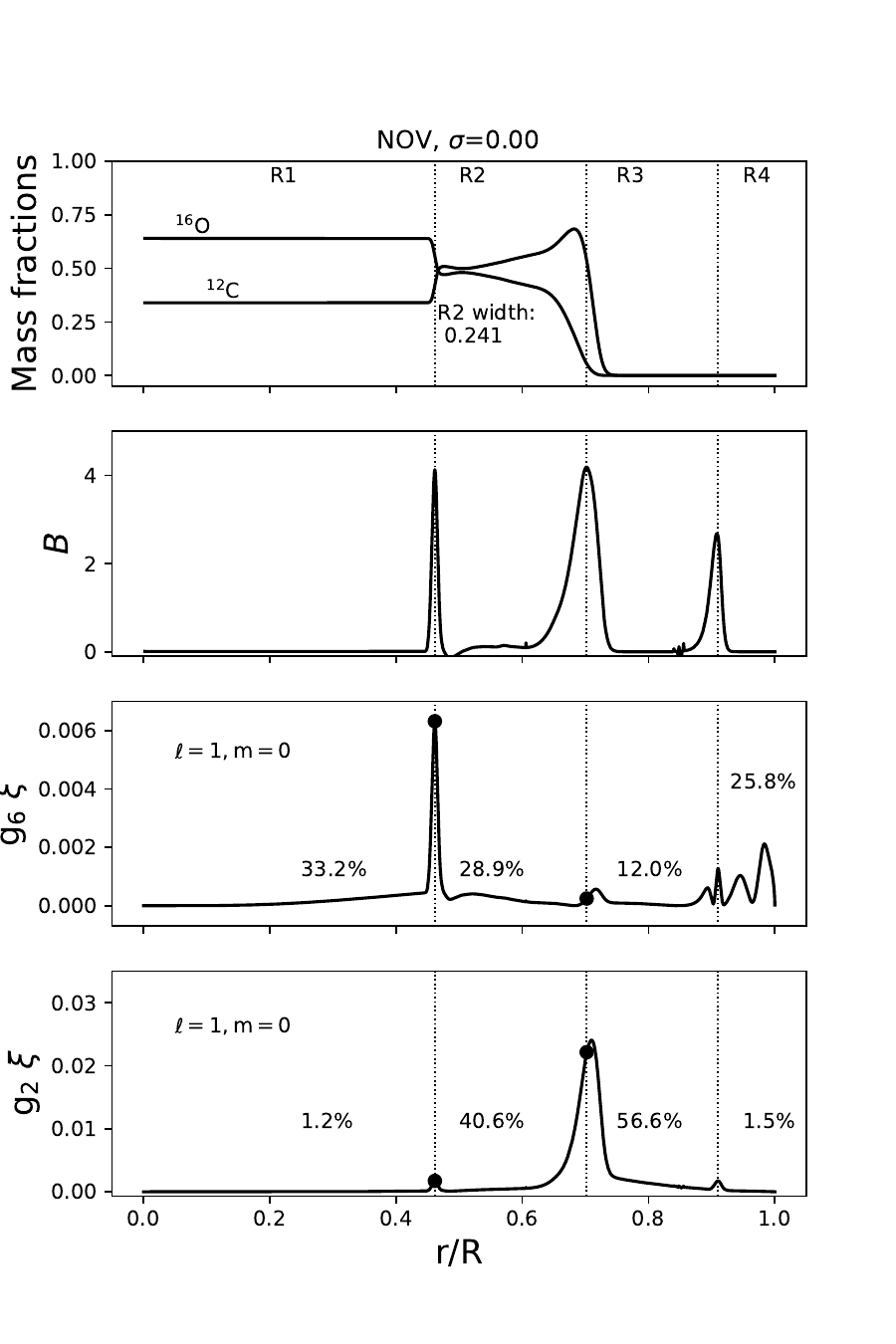}
\includegraphics[trim={0cm 1cm 1.5cm 1cm},clip,width=0.44\textwidth]{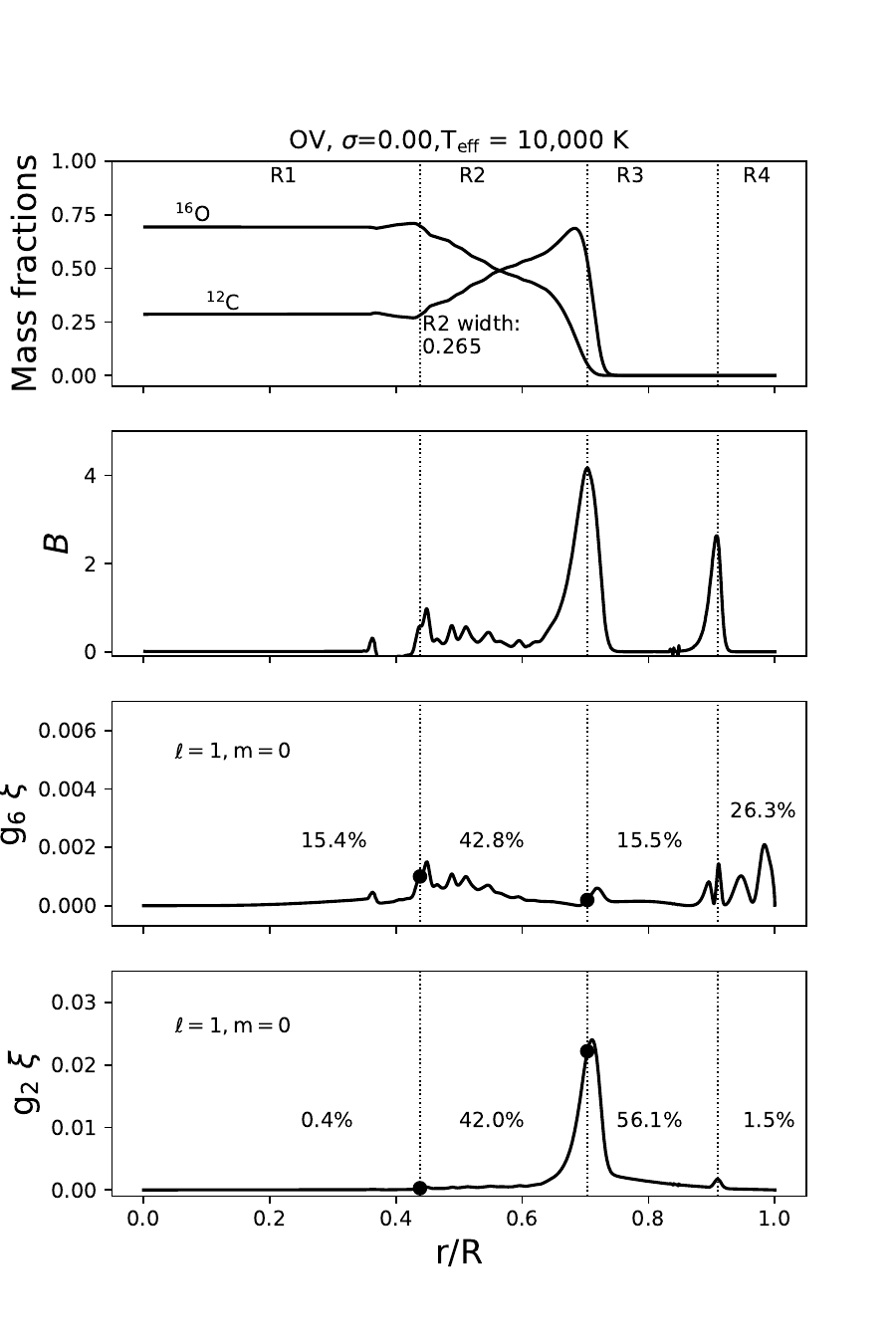}
\end{center}
\end{interactive}
\caption{\textit{Top to bottom:} Mass fractions of $^{12}$C and $^{16}$O; $B$ profile; normalized weight function profile for the g$_6$ mode; normalized weight function profile for the g$_2$mode.  
The left column shows the NOV results and the right column shows the OV results.
Both figures are for $\sigma=0.0$, $\Teff=10,000$ K. The R1-R4 region boundaries are indicated by dashed vertical lines.  An interactive figure is provided in the online version. Its functionality compares the NOV and OV diagrams, as structured in this figure, for any given $\sigma_i$ $\COrate$ reaction rate.  To use the interactive figure, click on the `sigma' slider at the top, and slide through indices 0 through 12. Sliding through each index provides the user the side by side comparison of the diagrams for the respective $\sigma_i$ $\COrate$ reaction rate used in the evolution. Index 0 starts at $\sigma_{-3.0}$ and index 12 ends at $\sigma_{3.0}$.  
}
\label{fig:4panels}

\end{figure*}

Figure~\ref{fig:4panels} shows, from top to bottom, the mass fraction profiles, $B$, and the $g_6$ and $g_2$ mode weight functions $\zeta$ for the final WDs at $\Teff=10,000$ K.  The left and right columns are the NOV and OV results respectively. Here, we show the comparison for $\sigma=0.0$, but an interactive figure link is provided in the online article to compare these properties for any $\sigma_i$.  For all $\sigma_i$, NOV/OV comparisons, the dotted vertical lines mark the region boundary locations in each panel.  This is useful to compare where the boundary locations are across multiple profile properties.  For instance, the R1-R2 boundary marks the C$\rightarrow$O transition region, the first most prominent peak in $B$, and the first peak-like features in $g_6 \ \zeta$ and $g_2 \ \zeta$ in the NOV case.   Comparing the OV column to the NOV column, we see the global impacts from overshooting.  Overall, prominent features in the NOV set are lessened in magnitude in the OV set.  The C$\rightarrow$O transition is more gradual, lessening the composition gradient at the defined boundary.  This remarkably impacts the shape of $B$.  The first prominent peak after max(O) is much less in magnitude for all $\sigma_i$, and is not the only outstanding peak near the boundary. There are now multiple, smaller peaks in $B$ and the $g_6 \ \zeta$ near the R1-R2 boundary as opposed to one. 
There are slight deviations between NOV and OV in these profiles for the R3 and R4 regions of the WD, but the R1$\rightarrow$R2 region in these profiles was affected most.

The $g_6 \ \zeta$ and $g_2 \ \zeta$ panels in Figure~\ref{fig:4panels} note the weight percentages per region in the WD.  This tells each region's contribution to the overall mode period (frequency).  An interesting result for all $\sigma_i$ is that both the $g_2$ and $g_6$ modes decrease the amount of weight in R1 when overshoot is included, and increase the amount of weight in R2.  There is also a slight decrease in the weight of R3 for $g_2$ for all $\sigma_i$ when overshoot is included.  These results are important.  The R2 region is the most reliable region in terms of extracting the $\sigma_i$ rate signature.  When overshoot is included, the R2 contribution to the overall pulsation modes in $g_2$ and $g_6$ are accentuated, implying that these modes more reliably distinguish $\sigma_i$ than the NOV set. A quantitative analysis of each region's weight percentage contribution per $\sigma_i$ is given for both sets in Table~\ref{tab:g2_table} and Table~\ref{tab:g6_table} for $g_2$  and $g_6$ respectively.  Overall, Table~\ref{tab:g2_table} shows that R2 and R3 are the most heavily weighted regions for $g_2$'s period.  $G_6$ has more equitable weight dispersed across regions, but the combined weight of R1 and R2 accounts for $\sim 50 \%$ of the $g_6$ period for any given model.  As identified in Figure~\ref{fig:wd_profiles} and Figure~\ref{fig:4panels}, R1 and R2 are the most impacted regions in this study. A g-mode with about half its weight from those regions may pick up the detailed differences more so than modes weighted more in outer regions.  This may explain why Figure~\ref{fig:raw_periods} shows a larger spread in the $g_6$ periods as this g-mode is likely picking up the R1 and R2 contributions to its period better than other g-modes. 

\begin{deluxetable}{|c|c|c|c|c|c|c|c|c|}[!htb]
  \tablecolumns{9}
  \tablecaption{$g_2$ Weight Function Percentages Per WD Region \label{tab:g2_table}
  }
  \tablehead{
    \multicolumn{1}{|c|}{}    & 
    \multicolumn{2}{c|}{R1} &
    \multicolumn{2}{c|}{R2} &
    \multicolumn{2}{c|}{R3} &
    \multicolumn{2}{c|}{R4} \\
    \multicolumn{1}{|c|}{$\sigma_i$} &
    \multicolumn{1}{c}{NOV} &
    \multicolumn{1}{c|}{OV} &
    \multicolumn{1}{c}{NOV} &
    \multicolumn{1}{c|}{OV} &
    \multicolumn{1}{c}{NOV} &
    \multicolumn{1}{c|}{OV} &
    \multicolumn{1}{c}{NOV} &
    \multicolumn{1}{c|}{OV} 
    }
\startdata
-3.0 & 0.91 & 0.75 & 40.6 & 41.3 & 57.0 & 56.4 & 1.47 & 1.47 \\
-2.5 & 1.14 & 0.99 & 40.2 & 44.2 & 57.2 & 52.9 & 1.43 & 1.94 \\
-2.0 & 1.05 & 0.52 & 40.2 & 41.1 & 57.2 & 56.9 & 1.54 & 1.53 \\
-1.5 & 1.18 & 0.53 & 39.5 & 41.7 & 57.9 & 56.2 & 1.50 & 1.50 \\
-1.0 & 1.16 & 0.27 & 40.4 & 41.5 & 56.9 & 56.8 & 1.48 & 1.46 \\
-0.5 & 1.15 & 0.18 & 38.8 & 42.1 & 58.6 & 56.3 & 1.43 & 1.49 \\
0.0  & 1.25 & 0.38 & 40.6 & 42.0 & 56.6 & 56.1 & 1.52 & 1.47 \\
0.5  & 1.44 & 0.49 & 40.8 & 41.9 & 56.2 & 56.2 & 1.52 & 1.47 \\
1.0  & 1.28 & 0.31 & 40.4 & 41.4 & 56.9 & 56.7 & 1.49 & 1.58 \\
1.5  & 1.32 & 0.28 & 39.9 & 41.4 & 57.2 & 56.8 & 1.50 & 1.51 \\
2.0  & 1.35 & 0.19 & 39.4 & 40.8 & 57.8 & 57.5 & 1.50 & 1.49 \\
2.5  & 1.25 & 0.42 & 38.3 & 41.6 & 58.9 & 56.6 & 1.47 & 1.45 \\
3.0  & 1.39 & 2.06 & 40.2 & 39.6 & 56.9 & 56.8 & 1.59 & 1.52 \\
\enddata
\end{deluxetable}

\begin{deluxetable}{|c|c|c|c|c|c|c|c|c|}[!htb]
  \tablecolumns{9}
  \tablecaption{$g_6$ Weight Function Percentages Per WD Region \label{tab:g6_table}
  }
  \tablehead{
    \multicolumn{1}{|c|}{}    & 
    \multicolumn{2}{c|}{R1} &
    \multicolumn{2}{c|}{R2} &
    \multicolumn{2}{c|}{R3} &
    \multicolumn{2}{c|}{R4} \\
    \multicolumn{1}{|c|}{$\sigma_i$} &
    \multicolumn{1}{c}{NOV} &
    \multicolumn{1}{c|}{OV} &
    \multicolumn{1}{c}{NOV} &
    \multicolumn{1}{c|}{OV} &
    \multicolumn{1}{c}{NOV} &
    \multicolumn{1}{c|}{OV} &
    \multicolumn{1}{c}{NOV} &
    \multicolumn{1}{c|}{OV} 
    }
\startdata
-3.0 & 25.5 & 20.1 & 25.6 & 32.4 & 21.1 & 19.8 & 27.8 & 27.8 \\
-2.5 & 33.1 & 19.1 & 29.5 & 33.5 & 13.1 & 20.2 & 24.2 & 24.2 \\
-2.0 & 32.3 & 16.6 & 30.8 & 36.3 & 13.9 & 19.7 & 23.0 & 23.0 \\ 
-1.5 & 33.5 & 17.3 & 29.6 & 39.1 & 12.6 & 17.3 & 24.4 & 24.4 \\ 
-1.0 & 33.8 & 13.4 & 30.0 & 43.1 & 12.9 & 17.4 & 23.3 & 23.3 \\  
-0.5 & 33.5 & 11.7 & 29.8 & 47.5 & 12.8 & 14.9 & 23.9 & 23.9 \\  
0.0  & 33.2 & 15.4 & 28.9 & 42.8 & 12.0 & 15.5 & 25.9 & 25.9 \\  
0.5  & 26.6 & 16.4 & 22.5 & 41.0 & 13.8 & 14.0 & 37.1 & 37.1 \\
1.0  & 31.2 & 14.1 & 27.1 & 43.8 & 12.4 & 16.1 & 29.3 & 29.3 \\ 
1.5  & 32.2 & 13.7 & 27.4 & 46.7 & 12.2 & 14.7 & 28.3 & 28.3 \\ 
2.0  & 25.5 & 11.7 & 23.0 & 48.1 & 14.1 & 14.3 & 37.3 & 37.3 \\ 
2.5  & 30.9 & 14.2 & 28.0 & 42.5 & 12.5 & 13.8 & 28.6 & 28.6 \\ 
3.0  & 30.1 & 32.0 & 25.5 & 26.2 & 12.4 & 13.8 & 32.0 & 32.0 \\ 
\enddata
\end{deluxetable}

\begin{figure*}[ht!]
\includegraphics[width=0.48\textwidth]{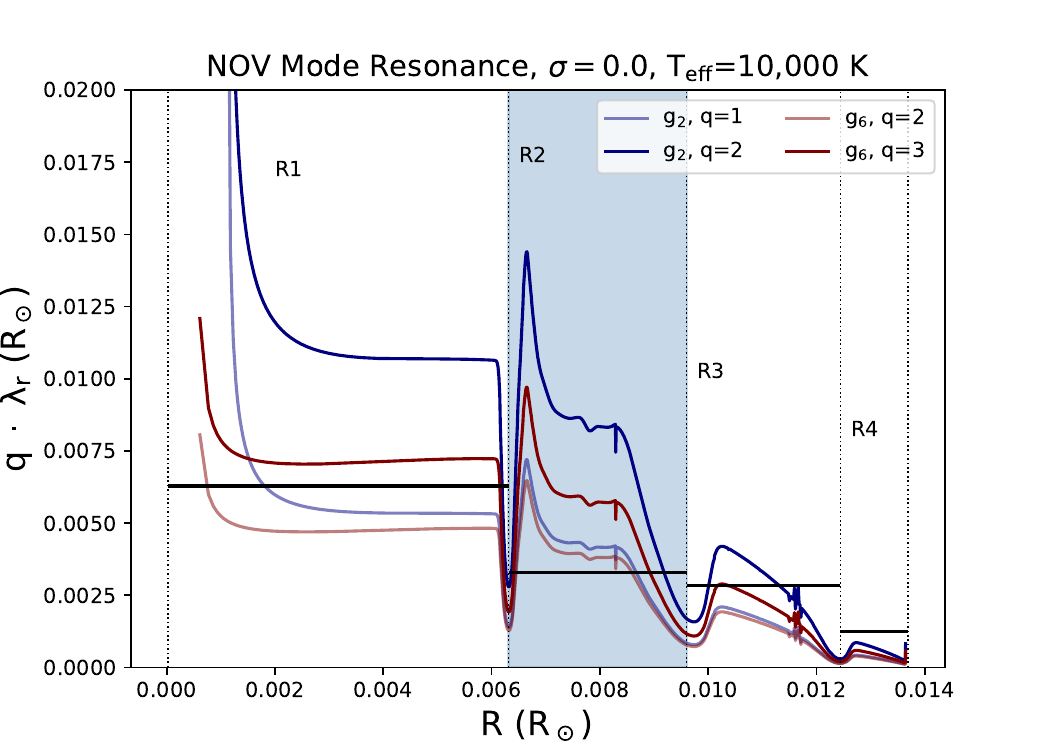}
\includegraphics[width=0.48\textwidth]{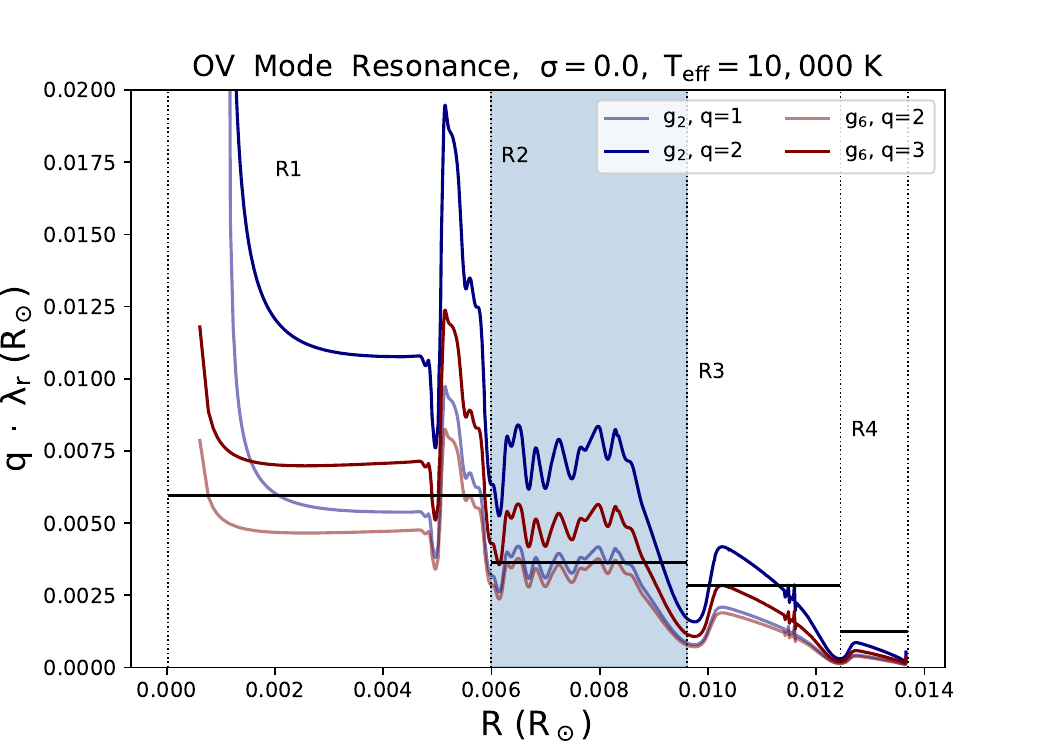}
\caption{Integer multiples of the local radial wavelengths $q \cdot \lambda_r$  for $g_2$ and $g_6$ as a function of the star's radius $R$.  Mode resonance occurs when $q \cdot \lambda_r$ closely matches the width of a certain region(s) in the star.  The left panel is the NOV set's mode resonance and the right panel is the OV set's mode resonance.  In both panels, the black horizontal lines mark the respective region widths.  Blue curves show $q \cdot \lambda_r$ for $g_2$ and maroon curves show $q \cdot \lambda_r$ for $g_6$. The $q$ values are stated in the legend.}
\label{fig:g-mode_resonance}
\end{figure*}

When an integer multiple $q$ of the local radial wavelength $\lambda_r$ for a given g-mode nearly matches the width of a certain region(s) in a star, the g-mode resonates with that region(s).  Figure~\ref{fig:g-mode_resonance} shows  $q\cdot\lambda_r (R_\odot)$ as a function of radius $R \ (R_\odot)$ for the $g_2$ and $g_6$ modes.  The NOV set doesn't show any particular close matches for any region.  But the closest matches to the R2 width were the $\lambda_r$ curves of $g_2, q=1$, and $g_6, q=2$.  Further, the $g_2, q=2$ and $g_6, q=3$ modes were best at resonating with R3.  Larger $q$ values may show stronger resonance with R4. The resonance with R2 is enhanced in the OV set.  The $g_2, q=1$ and $g_6, q=2$ $\lambda_r$ curves match much more closely to the R2 width in the OV set. This implies that overshoot has enhanced the g-mode resonance for our signature modes in the region that was constructed mainly from radiative burning (Figure~\ref{fig:kipps}).  We also see stronger resonance within the R1 region with the $g_2, q=1$ $\lambda_r$ curve.

\begin{figure*}[ht!]
    \includegraphics[trim={0cm 0cm 0cm 0.1cm},clip,width=0.48\textwidth]{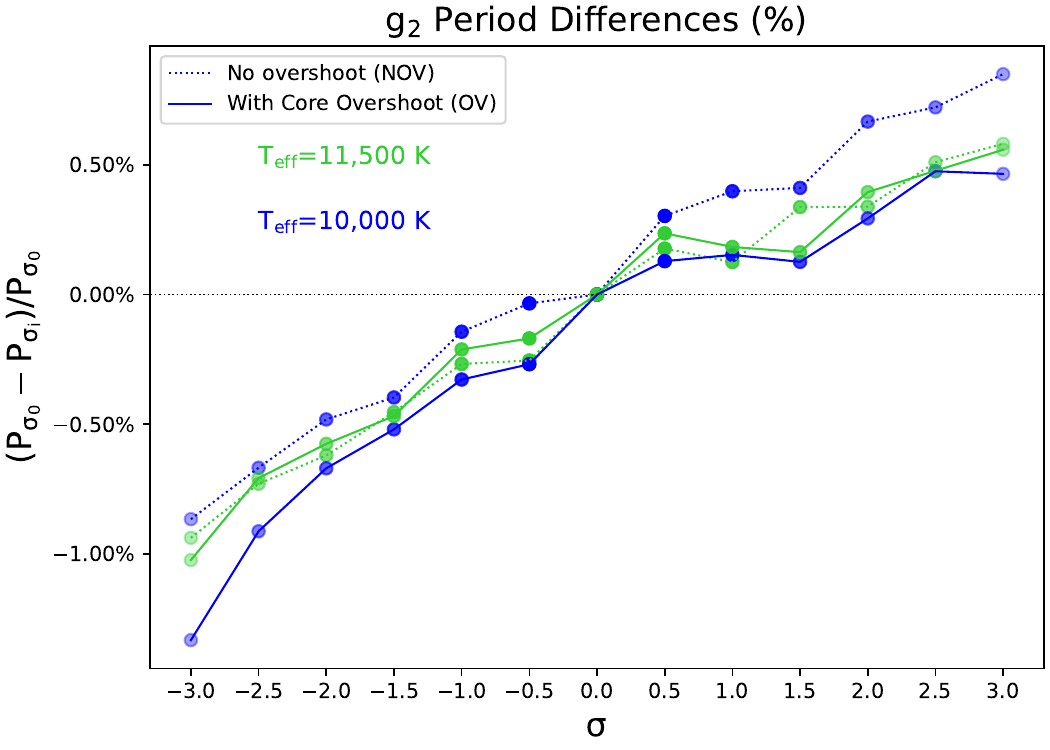}
   \includegraphics[width=0.48\textwidth]{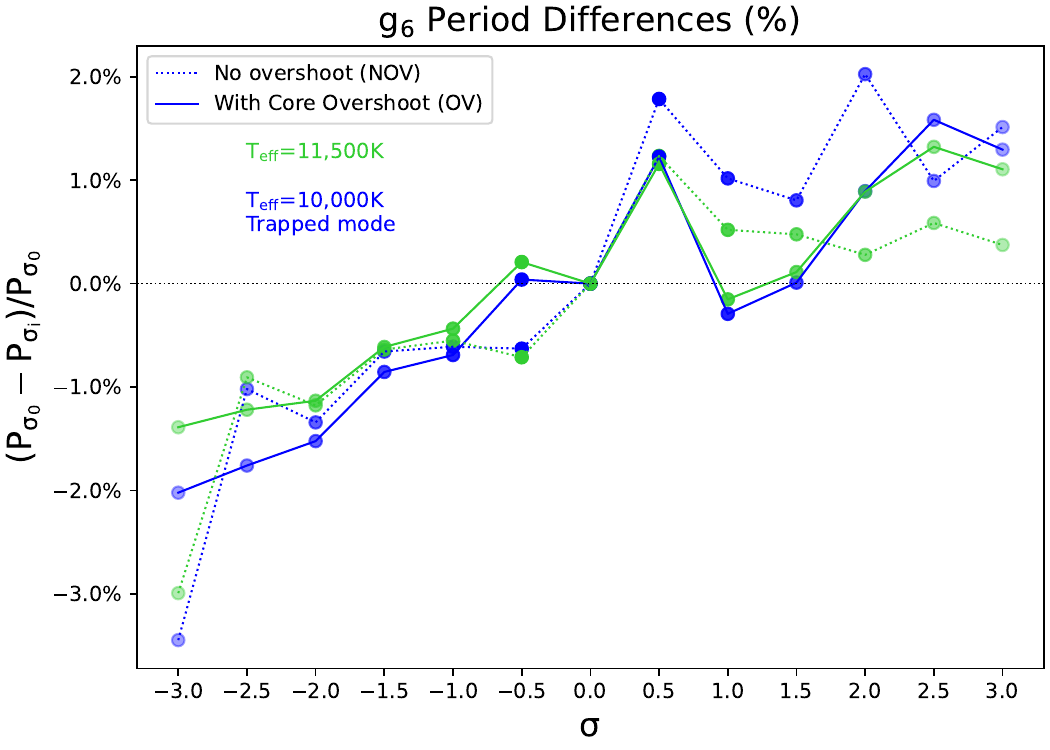}
    \caption{
 Adiabatic g$_2$ (left) and $g_6$ (right) mode signatures for the NOV and OV  sets, at $\Teff=11,500$ (bright green) and $\Teff=10,000$ K (blue) respectively}.
    \label{fig:pulsation_stuff}
\end{figure*}

Will the differences between the NOV and OV sets in Figure~\ref{fig:4panels} impact the WD $\COrate$ $\sigma_i$ pulsation signatures shown in Figure~\ref{fig:nov_signals}? Figure~\ref{fig:pulsation_stuff} shows the resulting relative period percent differences, as a function of $\sigma_i$ at $\Teff=11,500$~K (bright green) and $\Teff=10,000$~K (blue).  The period differences are negative for $\sigma_i$ with longer periods than the $\sigma=0$ model, and are positive for $\sigma_i$ with shorter periods than the $\sigma=0$ model for the given NOV or OV set.  The left of this figure shows the period differences for $g_2$, and the right shows the period differences for $g_6$.  The NOV set is indicated by the dotted lines and the OV set is the solid lines.  

Looking at $g_2$, the period differences between NOV and OV at $\Teff=11,500$ K are minimal; both sets show a trend of decreasing period with increasing $\sigma_i$.  At $\Teff=10,000$ K, the OV set shows an overall decrease in the percent differences, and a slightly greater variation in the overall $\sigma_i$ vs. $g_2$ period difference shape.  However, at both temperatures, the same pattern of the $g_2$ period decreasing with increasing $\sigma_i$ is sustained with overshoot inclusion.
Further, the magnitude of percent differences, ranging from $\simeq$\,$-$1.5 to +1.0, is within the detectable threshold \citep{chidester_2021_aa}.

The OV set shows greater deviation from the NOV line of period percent differences in $g_6$ more-so than $g_2$.  This is most likely because $g_6$ is more sensitive to changes from R1 than $g_2$.  Nonetheless, despite the $\sigma_{=-0.5}$ and $\sigma_{+1.0}$ outliers, the overall trend remains: $\sigma_i<0$ generally have longer periods than $\sigma_0$ and $\sigma_i>0$ generally have shorter periods than $\sigma_0$. Once again, the magnitude of the relative period percent differences surpass the observable threshold. 

An interesting note is that for both $g_2$ ad $g_6$ signals, the percent differences change more in the NOV set as the models cool from $\Teff=11,500$ to $\Teff=10,000$  K than the OV set.  The OV set showed nearly the same period differences at both temperatures.

\section{Discussion}
\label{sec:discussion}
C22 found pulsation signature(s) for the experimental $\COrate$ reaction rate probability distribution function. They describe four sensitivities that may impact this result: width of the O$\rightarrow$C transition, mixing during CHeB, thermal pulse history on the AGB, and the $3\alpha$ reaction rate.
This work investigated the impact that overshoot inclusion had on the $\COrate$ reaction rate pulsation signature(s).  Doing so, we address the width of the O$\rightarrow$C transition and mixing during CHeB.  Further, by ignoring the thermal pulse history in our models, we also address the sensitivity to the number of thermal pulses, albeit, the trivial case when the number of thermal pulses is zero.  In the following paragraphs, we discuss how these three sensitivities impacted our results.  We further caution how our results could be impacted from further sensitivity investigations.

Including overshooting overall increased the width of the O$\rightarrow$C transition for all $\sigma_i$ cool WDs.  This lessened the sharp peak in $B$ at the O$\rightarrow$C transition, and decreased the peak in $g_6 \ \zeta$ at the O$\rightarrow$C transition.  While the transition peak was lessened and dispersed into R2, widening the O$\rightarrow$C transition shows an \textit{enhancement} of both the weight contribution to the R2 region for $g_2$ and $g_6$, and the R2 resonance with $\lambda_r$ for $g_2$ and $g_6$.  The widening of the O$\rightarrow$C transition was from the combined effects of overshoot inclusion and the $\sigma_i$ prescription.  We conclude that widening the O$\rightarrow$C transition imposes differences in $B$, $\zeta$, and the pulsation periods. Despite these changes, we still find the $g_2$ and $g_6$ relative period differences in the NOV and OV sets to distinguish the \COrate\ reaction rate probability distribution function.  Namely, the pattern of decreasing period with increasing $\sigma_i$ persisted in both NOV and OV sets.  By itself, the inclusion of overshooting does not destroy the seismic signatures of the \COrate\ reaction rate  in our WD models -- which was the primary question of this study.

We caution that increasing (decreasing) the width of the O$\rightarrow$C transition in CO WD models could potentially yield different results.  Our CO WD models were informed from their evolution history, with the stated model parameters.  Thus, an increase (decrease) of the width of the O$\rightarrow$C transition may come from choosing different mixing processes, prescriptions and parameters, such as for convection and overshooting. A change  in the width of the O$\rightarrow$C transition may also come from mixing processes not considered in this study such as 
time-dependent convection \citep{jermyn_2023_aa}, rotationally induced mixing, semiconvection, thermohaline mixing, 
\revision{or first-order phase separations of the CO mixture \citep{bauer_2023_aa}.}

Ignoring the thermal pulse history gave an additional low-order adiabatic g-mode signature for $\sigma_i$, namely the $g_2$ signal. This signal was not found in C22, where the thermal pulse history was included.  Future studies on the thermal pulse phase of evolution with different temporal and spatial resolutions are needed to determine the sustainability of the $g_2$ signal as a probe for $\sigma_i$. 
Concurrently, future studies could also explore the interaction, if any, between the thermal pulses and overshooting during CHeB on the chemical profiles.

The CO cores of WDs are the result of the competition between $3\alpha$ and $\COrate$ during CHeB.  An experimental $3\alpha$ reaction rate probability distribution function, similar to the existing one for $\COrate$
\citep[][C22]{deboer_2017_aa, mehta_2022_aa,farag_2022_aa}, does not yet exist to our knowledge, although a probability distribution function could be constructed using the STARLIB reaction rate library \citep{sallaska_2013_aa, fields_2016_aa, fields_2018_aa}.
Future studies involving both reaction rate probability distribution functions could probe properties of DAV WD models in the $3\alpha$ rate - $\COrate$ rate plane. For example, the $3\alpha$ reaction rate is likely to slowly modulate the central $^{16}$O mass fraction at any \COrate\ reaction rate because $3\alpha$ controls the production of $^{12}$C. The \COrate\ reaction rate will likely modulate the central $^{16}$O mass fraction more strongly at any $3\alpha$ reaction rate. \revision{We speculate} that the radiative region R2 will exist in all such models. We also \revision{suspect} that all such models, whether terminated at the first thermal pulse or evolved through the thermal pulse phase, will show a trapped mode, with substantial trapping from R2, that best probes the \revision{$^{12}C(\alpha, \gamma)^{16}O$  burning reaction rates} (i.e. $g_6$ in this work, and see Figure 9 in C22).  We caution that the relative period shifts we find in this work from considering the \COrate\ probability distribution and overshooting may change when a 3$\alpha$ reaction rate probability distribution function is also considered.

\cite{de-geronimo_2017_aa} found that including overshooting impacted ensuing WD pulsations by $\sim 2-5$ s.    
Their results were independent of their $\COrate$ reaction rate uncertainty evaluation. We combined the effects of overshooting and the $\COrate$ reaction rate sensitivities in our pulsation analysis, and likewise find period differences of similar magnitudes.  Our $\COrate$ reaction rate analysis spanned the current experimental probability distribution function, which analyzed different rate values than those explored in \cite{de-geronimo_2017_aa}.  They concluded that the $\COrate$ uncertainty was less relevant than overshooting.  In this study, we find that the combined effects from overshooting and the $\COrate$ reaction rate probability distribution function yields remarkable differences in the structure of the CO WDs, and pulsation differences.  Despite these differences, we still find pulsation signatures for $\sigma_i$.

\section{Summary}
\label{sec:summary}

We conducted a search for signatures of the current
experimental $\COrate$ reaction rate probability distribution function in the pulsation periods of CO WD models with the inclusion of overshooting.  We found two signature adiabatic g-modes that show period differences with the reaction rate probability distribution function $\sigma_i$ trend regardless of whether or not overshoot is included.  We find a $g_2$ period difference signature is inversely proportional to $\sigma_i$.  Without overshoot, the $g_2$ relative period differences span $\pm$ 0.9\%.  With overshoot, the $g_2$ relative period differences range from -1.33\% to 0.47\%.  The average magnitude of the relative period differences for $g_2$ were 0.46\% and 0.44\% respectively.  The $g_6$ period differences were larger in magnitude, spanning from -3.44\% to 1.78\% for NOV and -2.02\% to 1.58\% for OV.  The average magnitude of  the $g_6$ period differences were 1.21\% and 0.95\% respectively.  The average magnitudes of the $g_2$ and $g_6$ period differences were slightly decreased from the NOV set.  

We found that the R2 weight contribution to these g-modes was enhanced with overshoot inclusion.  The R2 region remains the best identifying region for tracing the $\COrate$ reaction rate probability distribution function.  This is because even with overshoot inclusion, it is predominantly constructed by radiative burning during CHeB.  
Regardless of whether or not overshooting is considered, we find:
\begin{enumerate}
    \item two signature g-modes, $g_2$ and $g_6$ probe $\sigma_i$
    \item $g_2$ is inversely proportional to $\sigma_i$ and $g_6$ is a trapped mode
    \item the $g_2$ and $g_6$ periods are generally shorter for positive $\sigma_i$ and longer for negative $\sigma_i$
    \item both signatures have period deviations within the detectable regime
\end{enumerate}

These findings suggest that an astrophysical constraint on the $\COrate$ reaction rate probability distribution function remains, in principle,
extractable from the period spectrum of observed variable WDs.


\section{Acknowledgements}
We thank James Deboer for sharing the $^{12}$C$(\alpha,\gamma)^{16}$O probability
distribution function, Josiah Schwab for sharing \code{wd\_builder}, 
and Pablo Marchant for sharing \code{mkipp}.
We acknowledge using ChatGPT \citep{openai_2023_aa} to polish the language of one paragraph \citep{vishniac_2023_aa}.
This research is supported by NASA under the Astrophysics Theory Program grant NNH21ZDA001N-ATP, and in part by the National Science Foundation under Grant No. NSF PHY-1748958.
This research made extensive use of the SAO/NASA Astrophysics Data System (ADS).

\software{
\MESA\ \citep[][\url{https://docs.mesastar.org/}]{paxton_2011_aa,paxton_2013_aa,paxton_2015_aa,paxton_2018_aa,paxton_2019_aa,jermyn_2023_aa},
\texttt{MESASDK} 20190830 \citep{mesasdk_linux,mesasdk_macos},
\code{wd\_builder} \url{https://github.com/jschwab/wd_builder},
\GYRE\ \citep[][\url{ https://github.com/rhdtownsend/gyre}]{townsend_2013_aa,townsend_2018_aa},
\code{mkipp} \url{https://github.com/orlox/mkipp},
\texttt{matplotlib} \citep{hunter_2007_aa},
\texttt{NumPy} \citep{der_walt_2011_aa}, amd 
\texttt{ChatGPT} \citep{openai_2023_aa}.}

\appendix

\section{Microphpysics in MESA \label{app:A}}
The MESA EOS is a blend of the OPAL \citep{rogers2002ApJ...576.1064R}, SCVH
\citep{saumon_1995_aa}, FreeEOS \citep{irwin_2004_freeeos}, HELM \citep{timmes_2000_ab},
PC \citep{2010_potekhin}, and Skye \citep{jermyn_2022_aa} EOSes.

Radiative opacities are primarily from OPAL \citep{iglesias_1993ApJ,iglesias_1996_aa}, with low-temperature data from \citet{ferguson2005ApJ...623..585F}
and the high-temperature, Compton-scattering dominated regime by
\citet{poutanen2017ApJ...835..119P}.  Electron conduction opacities are from
\citet{cassisi_2007_aa} and \citet{blouin_2020ApJ...899...46B}.

Nuclear reaction rates are from JINA REACLIB \citep{cyburt_2010_ab}, NACRE \citep{angulo_1999_aa} and
additional tabulated weak reaction rates \citet{fuller_1985_aa, oda_1994_aa,langanke_2000_aa}.  Screening is included via the prescription of \citet{chugunov_2007_aa}.
Thermal neutrino loss rates are from \citet{itoh_1996_aa}.

\section{Model Optimization and Resolution \label{app:B}}
\subsection{Reduced Chemical Network}
\begin{figure*}[ht!]
    \centering
    \includegraphics[width=1.0\apjcolwidth]{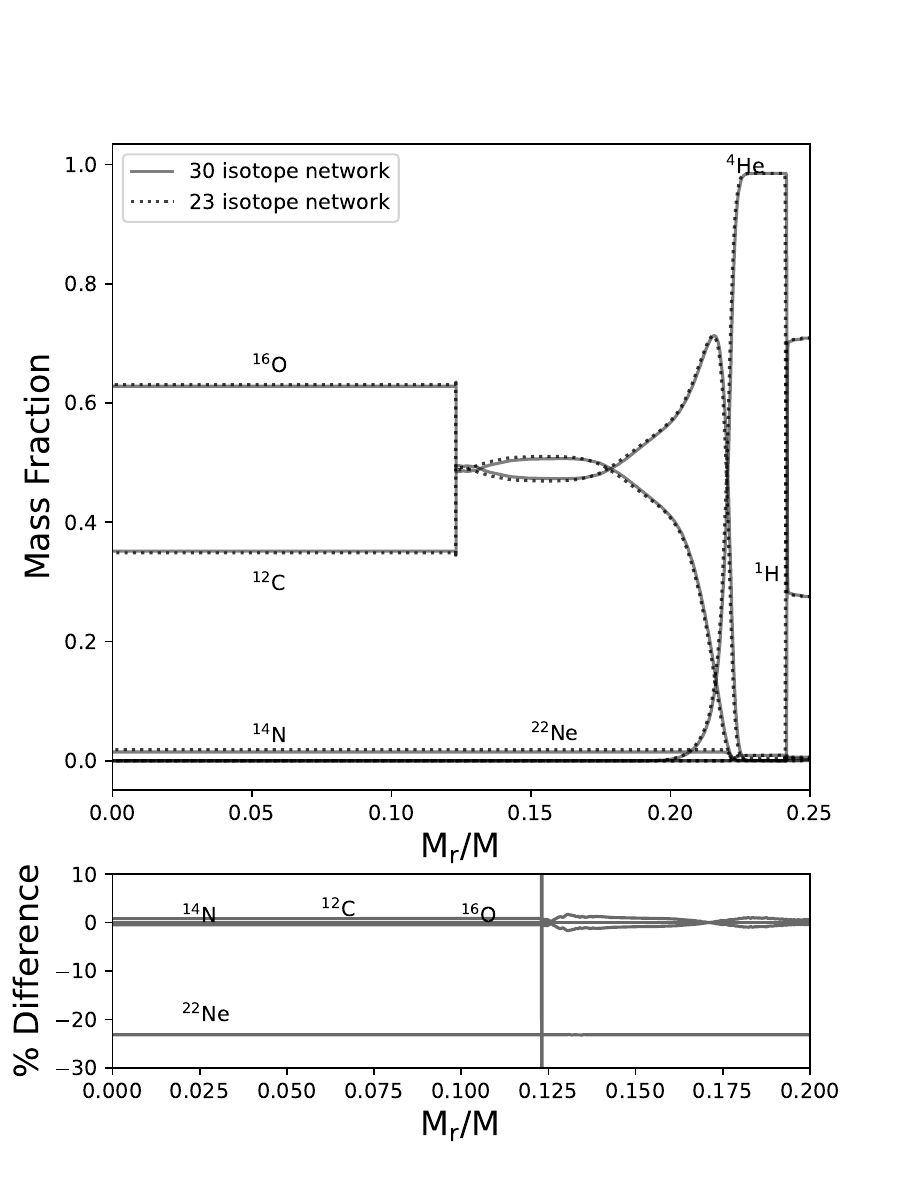}
    \caption{Mass fraction profiles for at the completion of CHeB for a 30 isotope (solid) and 23 isotope nuclear reaction network (dotted).  Shown are the 5 most abundant isotopes for both networks.}
    \label{fig:appendix1}
\end{figure*}

The nature of our evolutionary models is computationally expensive. This paper is concerned about overshooting and the $\COrate$ reaction rate probability distribution function, which primarily dictate the evolutionary processes and consequences of the CHeB phase.  The isotopes most impacted during CHeB are $\carbon$, $\oxygen$, and $\helium$. $\nitrogen$ and $\neon$ are the next two most impacted isotopes during CHeB.  We thus optimize the efficiency of our models by reducing the chemical network number of isotopes from 30 to 23.  The eliminated isotopes are $^{21}$Ne, $^{21,22,23}$Na, $^{23,24}$Mg, and $^{56}$Fe. A comparison of the resulting inner mass fraction profiles for the 5 most abundant isotopes for both networks is shown in Figure~\ref{fig:appendix1} for each chemical network.  This figure shows the profiles at the completion of CHeB.  both network models used the same temporal and spatial resolution during CHeB.  The run-time was reduced from a few days to a a few hours on 12 cores. All resolution studies were conducted with $\sigma=0.0$ without overshoot (NOV).  

Reducing the network impacted $\neon[22]$ most, with an offset of $\sim 22\%$ more $\neon[22]$ in the 23 isotope network.   We note that C22 used a 30 network and our overall signature results persistent through variations in heavier isotopes. 


\subsection{Temporal Resolution}

Several timestep limiters in $\MESA$ help optimize convergence studies.  In this paper, we want to limit the timestep to achieve the temporal resolution that yields a smooth evolution of the central  $\helium$, $\oxygen$, and $\carbon$ abundances during CHeB.   We first utilize the \code{delta\_XC\_cntr\_limit} limiter.  This limits the amount the central $\carbon$ abundance can change in a given timestep.  To help optimize computational run-time, we begin limiting the change in central $\carbon$ during CHeB which the central helium abundance X$(\helium_c)<0.6$.  This is done by adding the following lines of code in the $\MESA$ \code{run\_star\_extras.f90} file:\\

\begin{itemize}\tightitems
\item[] 
\texttt{if ((s\% center\_h1$<$1d-6).and.(s\% center\_he4 $<$ 0.6).and.(s\% delta\_XC\_cntr\_limit$>$0.001))then}
\item[]
\texttt{   s\% delta\_XC\_cntr\_limit = 0.0005d0}  
\item[]
\texttt{end if}
\end{itemize}

This temporal resolution was used for the 30 and 23 isotope network models. We refer to it as resolution A.  The remaining temporal resolution studies were performed using the 23 isotope chemical network.

The next iteration of increased temporal resolution modified the \code{run\_star\_extras.f90} file to include the following:\\
  \begin{itemize}\tightitems
  \item[]
  \texttt{if ((s\% center\_h1$<$1d-6).and.(s\% center\_he4 $<$ 0.5).and.(s\% delta\_XC\_cntr\_limit$>$0.001))then}
\item[]

   \texttt{          s\% delta\_XC\_cntr\_limit = 0.0005d0}
 \item[]
   \texttt{
            s\% delta\_lgT\_cntr\_limit = 5d-4}
  \item[]
  \texttt{
            s\% delta\_lgT\_cntr\_hard\_limit = 1d-3}
  \item[]
   \texttt{
            s\% delta\_lgRho\_cntr\_limit = 1d-3}
   \item[]
   \texttt{
            s\% delta\_lgRho\_cntr\_hard\_limit = 5d-3
 }
 \item[]
 \texttt{
end if}
\end{itemize}

This resolution is employed slightly earlier during CHeB, when X$(\helium_c)<0.5$.  We added limits to the change in central temperature and density from resolution A. This is resolution B.

Our third resolution iteration used the following limiter controls in the \code{run\_star\_extras.f90} file:\\
\begin{itemize}\tightitems
\item[]
  \texttt{if ((s\% center\_h1$<$1d-6).and.(s\% center\_he4 $<$ 0.999).and.(s\% delta\_XC\_cntr\_limit$>$0.001))then}
\item[]
  \texttt{  s\% delta\_XC\_cntr\_limit = 0.00025}
  \item[]
  \texttt{
            s\% delta\_XO\_cntr\_limit = 0.00025}
\item[]
  \texttt{
            s\% delta\_lgT\_cntr\_limit = 2.5d-4}
\item[]
  \texttt{
            s\% delta\_lgT\_cntr\_hard\_limit = 0.5d-3}
\item[]
  \texttt{
            s\% delta\_lgRho\_cntr\_limit = 0.5d-3}
\item[]
  \texttt{
            s\% delta\_lgRho\_cntr\_hard\_limit = 2.5d-3}
            \item[]
  \texttt{
    end if}
\end{itemize}

This is resolution C.  We have set the limiters at the start of CHeB, and have decreased the limiter values from those in resolution B. 

\begin{figure}[!htb]
    \centering
    \includegraphics[trim={0.5cm 0.cm 1.5cm 0cm},clip,width=0.32\textwidth]{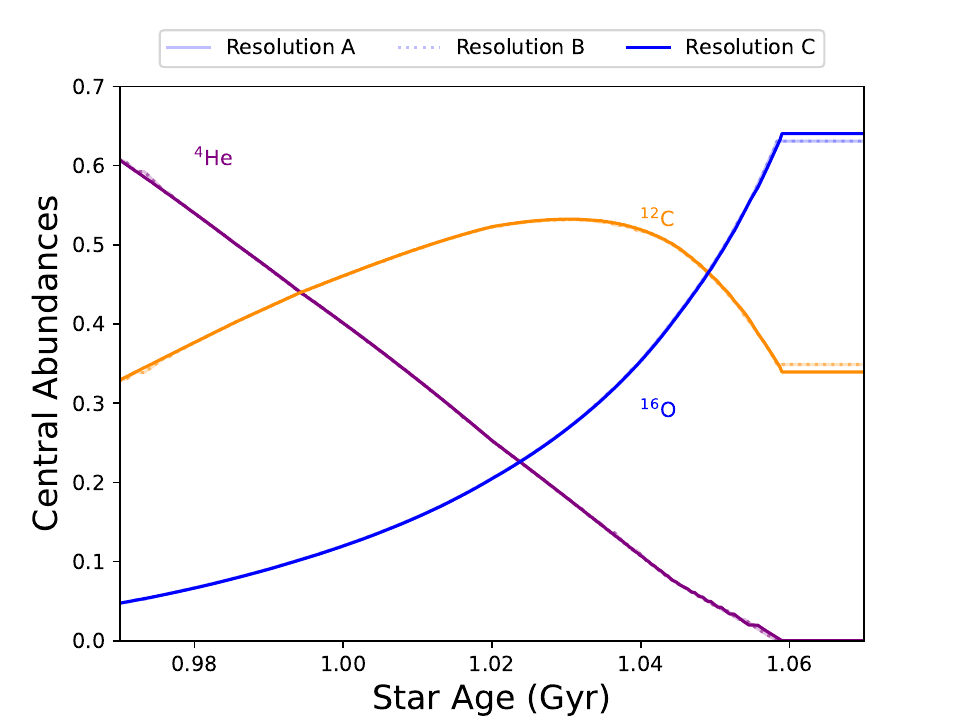}
    \includegraphics[trim={0.5cm 0.cm 1.5cm 1.3cm},clip,width=0.32\textwidth]{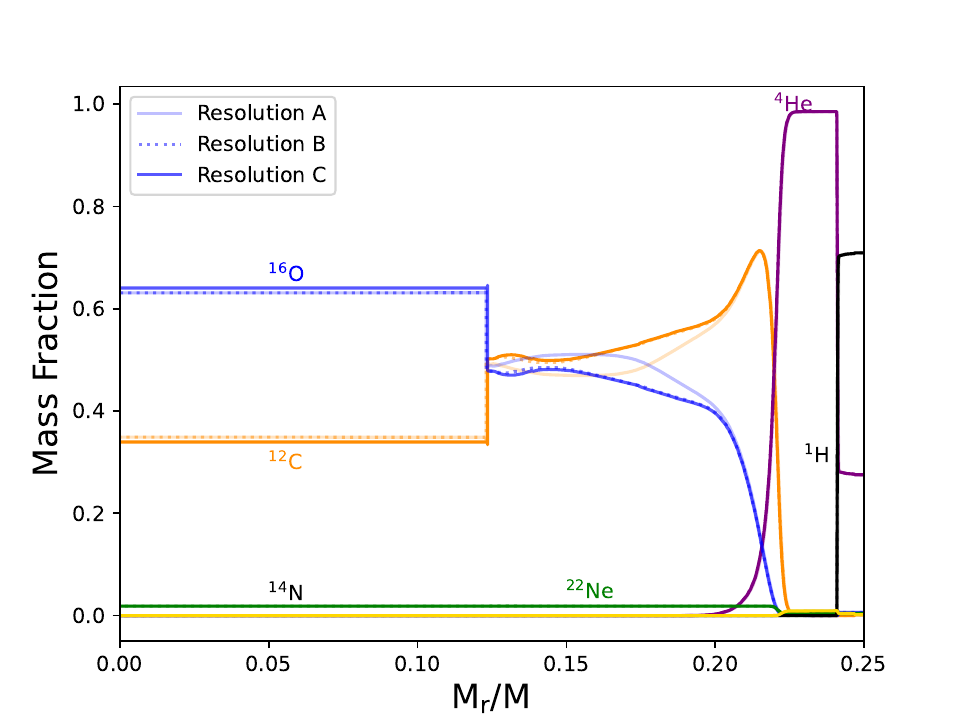}
    \includegraphics[trim={0.5cm 0.cm 1.6cm 1.3cm},clip,width=0.32\textwidth]{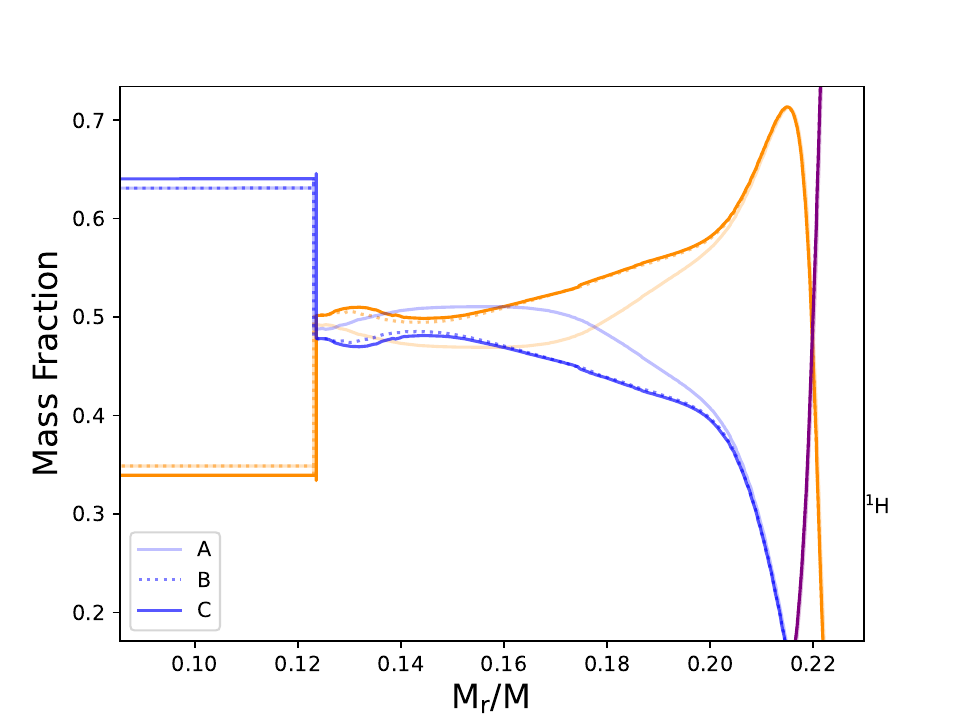}
    \caption{\textit{Left:} Evolution of central mass fractions during CHeB until $\log(L/\Lsun)$\,=\,3.0 at three resolutions.  \textit{Middle:} Mass fraction profiles at $\log(L/\Lsun)$\,=\,3.0 for the three resolutions. \textit{Right:} Middle figure zoom-in to show differences.}
    \label{fig:resolutionsABC}
\end{figure}

A comparison for resolutions A, B, and C are shown in Figure~\ref{fig:resolutionsABC}. In each column, the solid light curves represent resolution A, the dotted curves B, and the dark solid curves C.  

The left figure shows the evolution of central abundances of $\helium$, $\carbon$, and $\oxygen$ during CHeB, starting when X$(\helium_c)\lesssim0.6$ until the completion of CHeB.  The central abundances for resolutions A and B are nearly identical.  Resolution C varies slightly, with the final central $\oxygen$ abundance reaching a slightly larger amount than resolutions A and B.  Further, all three resolutions show a smooth evolution of these central abundances throughout CHeB.

The middle plot in Figure~\ref{fig:resolutionsABC} shows the mass fraction profiles at the completion of CHeB.  We show the 5 most abundant isotope profiles for each resolution.  The $\carbon$ and $\oxygen$ profiles for A are noticeably different than the profiles for B and C, especially after the O$\rightarrow$C transition.  This is more apparent in the right plot of Figure~\ref{fig:resolutionsABC}, which zooms in on the $\oxygen$ and $\carbon$ profiles of the three resolutions.  Resolution B follows A in the core, but then more closely aligns with C after the  O$\rightarrow$C transition.  Since resolutions B and C agree well, with only a slight difference in the central $\carbon$ and $\oxygen$ abundance, we set resolution C as the standard temporal resolution for our 13 models.  
 


\bibliographystyle{aasjournal}
\bibliography{paper}
\end{document}